\documentclass[twocolumn,aps,prb,twoside]{revtex4-2}

\usepackage[hidelinks]{hyperref}

\usepackage{amsmath,amssymb}
\usepackage{breqn}
\usepackage{graphicx}

\usepackage{siunitx}
\makeatletter
\let\cat@comma@active\@empty
\makeatother

\usepackage[usenames,dvipsnames]{xcolor}
\definecolor{TTH-color}{rgb}{0.11,0.66,0.11}

\definecolor{AMS-color}{rgb}{1,0.0,0.0}

\newcommand{\cc}{\text{c.~c.}}
\newcommand\kl[2]{{{#1}\,{#2}}} %
\newcommand{\mean}[1]{\langle{#1}\rangle}
\newcommand{\vib}{\text v}
\newcommand{\Ze}{\text{Ze}}
\newcommand{\dm}{\text{dm}}
\newcommand{\me}{\text{me}}
\newcommand{\mg}{\text{mg}}
\newcommand{\dr}{\text{d}}
\newcommand{\FMR}{\text{FMR}}
\newcommand{\Ms}{M_{\text s}}
\newcommand{\eff}{\text{eff}}
\newcommand{\ext}{\text{ext}}
\newcommand{\ZPF}{\text{ZPF}}
\newcommand{\inter}{\text{i}}

\begin{document}

\title{Signatures and characterization of dominating Kerr nonlinearity between \mbox{two driven systems} with application to a suspended magnetic beam}
\author{Andrii~M.~Sokolov}
\email[E-mail: ]{andriy145@gmail.com}
\affiliation{Department of Physics and Nanoscience Center, University of Jyvaskyla, P.O.~Box 35 (YFL), FI-40014 University of Jyv\"askyl\"a, Finland}
\affiliation{Institute of Physics of the National Academy of Sciences, pr.~Nauky 46, Kyiv 03028, Ukraine}
\author{Tero~T.~Heikkil\"a}
\affiliation{Department of Physics and Nanoscience Center, University of Jyvaskyla, P.O.~Box 35 (YFL), FI-40014 University of Jyv\"askyl\"a, Finland}

\begin{abstract}
We consider a model of two harmonically driven damped harmonic oscillators that are coupled linearly and with a cross-Kerr coupling.
We show how to distinguish this combination of coupling types from the case where a coupling of optomechanical type is present.
This can be useful for the characterization of various nonlinear systems, such as mechanical oscillators, qubits, and hybrid systems.
We then consider a hybrid system with linear and cross-Kerr interactions and a relatively high damping in one of the modes.
We derive a quantum Hamiltonian of a doubly clamped magnetic beam, showing that the cross-Kerr coupling is prominent there.
We discuss, in the classical limit, measurements of its linear response as well as the specific higher-harmonic responses.
These frequency-domain measurements can allow estimating the magnitude of the cross-Kerr coupling or the magnon population.
\end{abstract}

\maketitle

\section{Introduction}
\label{secIntro}

Kerr effect in optics arises due to the dependence of the electric displacement field on the third power of the electric field~\cite{landau1984electrodynamics}.
Often it is masked by the analogous Pockels effect quadratic in the electric field.
However, if the media is approximately inversion symmetric, the Pockels effect vanishes and the Kerr effect is the dominant nonlinearity~\cite{landau1984electrodynamics,melnichuk2010directkerr}.
In that case the Kerr effect also provides the dominating nonlinear interaction between different radiation modes, known as the cross-Kerr interaction~\cite{stolen1973kerrwaveguide}.
Similar nonlinearities appear in a plethora of other systems, when dynamical equations contain non-negligible terms cubic in the system variables.
Many of these systems are considered relevant to quantum technologies.
Transmon qubits~\cite{rasmussen2021sccompanion,koch2007transmon} naturally couple with a cross-Kerr interaction in the charge basis~\footnote{The respective interaction is of the same form as in the closely related Cooper pair box~\cite{rasmussen2021sccompanion}, for which the cross-Kerr-like form of the capacitive interaction is elucidated in Ref.~\cite{wendin2007qubits}.}, while in the energy-eigenstate basis (which is usually used as the computational basis) the cross-Kerr coupling can be engineered with a SQUID coupler~\cite{neumeier2013sptransistor} or similar nonlinear circuits~(see Sec.~4 of Ref.~\cite{ciani2019hamiltoncompcpbcrosskerr} and references therein).
Dispersive interaction of a superconducting qubit and resonator photons, which is of a cross-Kerr form in the effective Hamiltonian~\cite{blais2004cavity,carbonaro1979canonical,sokolov2020singlephoton}, is routinely used for fast high-fidelity quantum non-demolition readout of a qubit~\cite{walter2017rapid,heinsoo2018rapid} and can be used for the quantum non-demolition measurement of the radiation state~\cite{johnson2010qnd}.

Interfacing quantum systems of different nature can combine their strengths for applications in quantum technologies~\cite{kurizki2015quantum}.
In a hybrid system, its parts of different nature interact despite having mismatched characteristic frequencies.
Often one part of a system has a much lower resonance linewidth than the other one.
That is usually the case for magnomechanics, where a high-Q mechanical mode interacts with a relatively lossy ferromagnetic resonance mode~\cite{zhang2016cavity,potts2021dynamical,kansanen2021magnomechanics,shen2022bistability}.
In this case the effect of coupling to the high-Q mode can be accessed by looking for narrow linewidth (first or second) sidebands in the response of the other mode, arising from nonlinear interactions between the two.
As we discuss in this paper, the position and strength of such sidebands depends on the type of nonlinearity.

\begin{figure}
\includegraphics{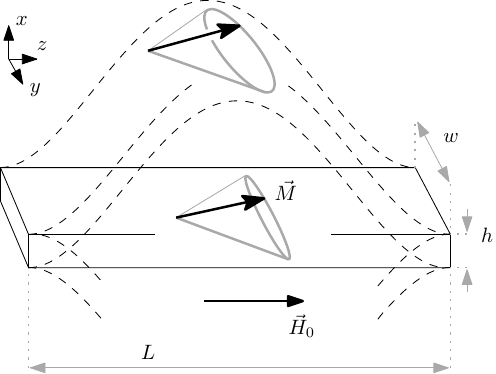}
\caption{Doubly clamped magnetic beam where the ferromagnetic resonance (FMR) mode interacts with the beam mechanical mode.
	The beam is long in the direction between its fixed points and has a wide cross-section, i.e.\ $L \gg w \gg h$.
	A symmetrical bending mode of the beam is excited.
	We assume that the beam is uniformly magnetized.
	A static magnetic field $\vec H_0$ is applied along the beam.
	FMR arises when an applied rf-field makes the beam magnetization $\vec M$ to precess around $\vec H_0$.
	Deformation of the beam changes the shape and the intrinsic anisotropies of the magnetic medium, which results in an interaction between the mechanical and the FMR mode.}
\label{figBeam}
\end{figure}

In magnomechanical systems, the third-order cross-Kerr interaction can dominate over the second-order coupling of optomechanical form if the latter is prohibited by the system symmetry. { This is an opposite regime than the one realized in Refs.~\cite{zhang2016cavity,potts2021dynamical,shen2022bistability}.}
For example, in Ref.~\cite{kansanen2021magnomechanics} with one of the authors, we predict the presence of a second-order nonlinearity in the equations of motion---the ``optomechanical'' coupling---arising due to interaction of vibrations of a doubly clamped magnetic beam and its macrospin mode.
However, to engineer this coupling, a static deformation of the beam is required.
The optomechanical coupling term in the Hamiltonian and the respective energy are of the first order in the beam displacement in its bending mode~(see Fig.~\ref{figBeam}).
The optomechanical energy changes its sign with respect to the displacement if the magnetic-related factors in it stay invariant.
In that case, the respective Hamiltonian term is zero if the beam possesses an inversion symmetry.
Similarly, if an optomechanical setup~\cite{kurizki2015quantum} possesses an inversion symmetry, the dominant nonlinearity there is of a cross-Kerr type~\cite{thompson2008strong,sankey2010strong,xuereb2013selectable,gong2009effkerr}.
In such cases, it is interesting to experimentally determine the form of the dominating nonlinear interaction.
As we discuss in this paper, this can be done by considering the nonlinear response of the two systems.

While often one is interested in quantum effects in these systems, they are not always easy to study.
For example, to maintain full quantum coherence of a magnomechanical system, thermal phonons should not randomly excite the mechanical oscillator.
Usually that means that the system has to be cooled down close to the ground state of mechanical vibrations, in practice to the temperature of a few millikelvin~\cite{aspelmeyer2014optomechanics,massel2011microwave}.
Theory wise, even linear interactions are not trivial to study in quantum systems~\cite{sokolov2020singlephoton,sokolov2020superconducting}.
Study of a nonlinear quantum coupling is even more challenging technically---for example, see Ref.~\cite{felicetti2018ultrastrong}, where it required performing high-order expansions in non-commuting operators and various approximations to calculate the system response.
However, we expect that the classical limit can be an insightful first step in understanding and characterizing the cross-Kerr interaction.

Two classical systems with the $\propto(x - x')^4$ cross-Kerr term in the Hamiltonian, where $x$ and $x'$ are each system coordinates, were studied in Ref.~\cite{musielak2005chaos}.
This form is natural for mechanical systems, and is the form that is engineered for transmon qubits in the computational basis~(see the references above).
However, in magnomechanics the interaction term can be $\propto x^2 x^{\prime 2}$.
When the processes involving several excitations from one system can be neglected, both forms provide the same resonant contributions to the population.
However, they generate different higher-order harmonics.
The common rotating-wave approximation~(RWA) analysis of the cross-Kerr coupling~\cite{massel2015ckoptmech,xiong2016ckoptmech,shen2022bistability,neumeier2013sptransistor,ciani2019hamiltoncompcpbcrosskerr,bruschi2020Uck,khorasani2019solutionckoptm} neglects the difference between the two forms.

Still, some properties of cross-Kerr coupling of the form $\propto x^2 x^{\prime 2}$ that arises in magnomechanics were considered in the literature.
The general form of a nonlinear coupling of the third order in the equations of motion was studied in Ref.~\cite{clementi2020internalresonance}.
The cross-Kerr coupling of magnomechanical form was considered in Refs.~\cite{manevitch2003free,givois2020backbone,hill2015interpreting}.
The references~\cite{clementi2020internalresonance,manevitch2003free,givois2020backbone} study the case when the linear resonances of the coupled systems are close to each other and to the drive frequency.
The case of 1:2 ratio of the linear resonances of the coupled systems was considered in Ref.~\cite{shami2022nonlinear12} for the same type of coupling.
In the mentioned works only a single tone drives the systems.
However, we expect that the use of two distinct tones can provide useful information in the system response.
The work~\cite{givois2020backbone} stands out from the rest of the cited above: there, the model parameters were fitted to the experimental response curves of a mechanical system.

In our contribution, we also seek how to characterize a system coupling.
First of our main results is general enough for any type of self- and cross-Kerr couplings:
We consider how to differentiate a cross-Kerr coupling from an optomechanical one.
Then we further focus on the cross-Kerr coupling in a magnomechanical system that lacks an optomechanical coupling due to its symmetries.
In the cited works~\cite{manevitch2003free,givois2020backbone,hill2015interpreting,shami2022nonlinear12}, a great deal of attention is given to the multistability that arises in the coupled system.
Damping in the FMR mode is expected to exceed the nonlinearities in the system, which allows us to neglect multistability.
That simplifies our theory and the interpretation of the responses.
This regime can be relevant for other hybrid systems.
We provide a theory that allows for characterizing the cross-Kerr coupling from its nonlinear frequency responses.

In Sec.~\ref{secModel}, we present the general model considered in the paper and discuss the assumptions and approximations we use;
in particular, we discuss the validity of the classical approximation in Sec.~\ref{secClassicalLimit}.
Section \ref{secExpansion} presents the approach we use to calculate the harmonic structure and their weights.
In Sec.~\ref{secNo11}, we elucidate how the presence or absence of different harmonics generated in the system point to different types of interaction.
In Sec.~\ref{secNonlinSdb}, we focus on a magnomechanical system with dominating cross-Kerr nonlinearity: a symmetric doubly clamped magnetic beam.
We provide analytical expressions for the second sidebands, which are the lowest-order harmonics present in that system.
Based on those expressions, we discuss a spectroscopic method to assess the strength of the cross-Kerr coupling or the magnon population in Sec.~\ref{secMeasurement}.
We discuss hybridization of the mechanical mode harmonics in Sec.~\ref{secAnticrossing}.
That provides another spectroscopic method to assess the cross-Kerr coupling strength or the magnon population.
In Sec.~\ref{secConclusion}, we sum up the results and discuss the directions for future work.
Derivation of the Hamiltonian of the doubly clamped magnetic beam is delegated to Appendix~\ref{apMagnomechanicsDerivation}.
While in the main part of the paper we work in the classical limit, the Hamiltonian derived there is quantum.
We also provide estimates for the cross-Kerr coupling strength in the appendix.

\section{Model and assumptions}
\label{secModel}

To be specific, we consider the model Hamiltonian
\begin{align}
\label{eqH}
	H = {}&\omega_1 a^* a + \omega_2 b^* b
\nonumber
\\
		&{}+ f_1 (a e^{i\Omega_1 t} + \cc)
		   + f_2 (b e^{i\Omega_2 t} + \cc)
\nonumber
\\
		&{}+ g (a^* + a)(b^* + b) + {\frac 1 4} K (a^* + a)^2 (b^* + b)^2.
\end{align}
It is motivated by the Hamiltonian of the magnomechanical system in Fig.~\ref{figBeam} that we derive in Appendix~\ref{apMagnomechanicsDerivation}.
In the nomenclature of quantum theory, $a$ and $b$ are the annihilation operators of the oscillators with $\hbar = 1$.
However, as we work in the classical limit, $a$ and $b$ are merely complex numbers (hence the complex conjugates in the Hamiltonian).
If the oscillators are, for example, linear LC-tanks, these complex numbers are proportional to the phasors of either the capacitor voltage or the current through the inductance.
In magnomechanics, ${\rm Re}(b)$ and ${\rm Im}(b)$ denote the amplitudes of the precessing magnetization modes in the two directions orthogonal to the static magnetization (see Appendix~\ref{apMagnomechanicsDerivation} for details).
Moreover,
\begin{equation}
\label{eqaDef}
	a = \frac 1 {\sqrt 2} (\sqrt{m\omega_1} x - i \frac p {\sqrt{m\omega_1}}),
\end{equation}
where $x$ is the deformation amplitude of the beam, $m$ is its mass, and $p$ is the respective momentum.
In the main part of the paper, we operate with mean values and omit the averaging operation $\mean{\ldots}$ except when that can lead to a confusion.
Hamiltonian~\eqref{eqH}---with the energy measured in the frequency units---can be obtained in a purely classical manner with the definition~\eqref{eqaDef}, its conjugate, and the similar equations for $b$.
This is analogous to the treatment of Ref.~\cite{zakharov1985hamiltonian}.
The oscillators with bare frequencies $\omega_1$ and $\omega_2$ are driven harmonically with the respective frequencies $\Omega_1$ and $\Omega_2$.
In the Hamiltonian~\eqref{eqH}, we denote the magnitude of the linear coupling by $g$ and that of the cross-Kerr coupling by $K$.
The cross-Kerr coupling here is of the form that arises due to vibration-induced anisotropy in a magnet far from saturation.
We derive this term in Appendix~\ref{apMagnomechanicsDerivation} for a doubly clamped magnetic beam (see Fig.~\ref{figBeam}) neglecting its crystal anisotropy.
Note that in Section~\ref{secNo11} we consider a more general form of the Kerr nonlinearity.

The derivation of the beam Hamiltonian in Appendix~\ref{apMagnomechanicsDerivation} yields no coupling apart from the cross-Kerr one.
In Ref.~\cite{kansanen2021magnomechanics}, a linear and a coupling of optomechanical form quadratic in the magnetization were obtained for an initially buckled beam.
As discussed in the introduction, in our case the symmetry of an initially unbuckled beam precludes such an optomechanical coupling.
Judging from its form, the linear coupling from the reference vanishes due to the symmetries when the static magnetic field is oriented along the beam.
The reference provides only an optomechanical coupling that is quadratic in the magnetic system operators.
We also expect that the absence of an optomechanical coupling \emph{quadratic in the mechanical coordinate} is a consequence of the symmetries in the initially unbuckled beam.
With all that said, we still include the linear coupling in Hamiltonian~\eqref{eqH}:
We expect that of the interactions discussed above, the linear one is the most prominent to arise if the symmetries are slightly broken in a real beam.
Moreover, a linear interaction can be prominent in other systems.

We use the RWA for the drive term and do not use it for the coupling between the oscillators.
We assume that the driving is sufficiently weak and $f_1, f_2 \ll \Omega_1, \Omega_2$; and on the other hand it is not detuned too much~\cite{sokolov2020singlephoton} so that
\begin{equation}
\label{eqRWAdetuning}
	|\Omega_1 - \omega_1| \ll \Omega_1 + \omega_1,
\quad
	|\Omega_2 - \omega_2| \ll \Omega_2 + \omega_2.
\end{equation}
This assumption allows us to neglect all non-resonant contributions from the co-rotating terms in the drive part of the Hamiltonian (note that this approximation is not used in Sec.~\ref{secNo11}).
However, we are interested in the higher harmonics generated by the interaction between the oscillators.
They are off-resonant and have the order of magnitude $K/\Omega_{1,2}$ or of higher powers in $g/\Omega_{1,2}$ and $K/\Omega_{1,2}$.
Hence we do not use the RWA in the linear and the cross-Kerr coupling terms in the Hamiltonian~\eqref{eqH}.

Let us comment on how the Hamiltonian~\eqref{eqH} can be extended with the bath terms to describe dissipation~\cite{scully1997quantum,berman2011influence,sokolov2020singlephoton}.
There are co-rotating terms in the interaction part of the Hamiltonian: in that case one cannot use the RWA for the bath coupling in quantum systems~\cite{beadoin2011dissipation,bambaogawa2014recipe}.
As $a$ and $a^\dagger$ and $b$ and $b^\dagger$ do not commute in the quantum case, the ground state of the interacting systems is different from their individual ground states.
An RWA coupling to the baths steers the system towards the state where both subsystems are in their individual ground states.
As the respective combined state is not even a system eigenstate, this generates excitations that are radiated to the bath infinitely, which is unphysical.
However, this problem does not arise in the classical limit we are interested in.
In that case, both the full and the RWA form of the system-bath coupling steer the system to the equilibrium position with well-defined values of momenta and coordinates.

\subsection{Validity of the classical approach}
\label{secClassicalLimit}

Let us discuss when the classical limit is applicable.
In this work, we only follow the evolution of the mean values of the system coordinates.
According to the Ehrenfest theorem, the mean values evolve according to the classical theory only when the forces governing the evolution are expressed in terms of the averaged coordinates.
However, that is not the case for nonlinear systems.
In our case, a cross-Kerr term with $\mean{ba^*a}$ arises in the equation of motion for $\mean b$ along the similar ones.
Quantum coherence between the system parts, as well as its superposed states prevent the correlator to break as $\mean{ba^*a} = \mean b \mean{a^*} \mean a$.
In what follows, we seek the conditions for the quantum coherence to decay, which guarantees the validity of the classical limit.
Afterwards, we evaluate these conditions for the magnomechanical system of the CoFeB doubly clamped beam.

First, quantum coherence decays due to random absorption and emission of excitations to the baths.
These processes~\cite{clerk2010introduction} occur with the rate
\begin{equation}
\label{eqDecoherenceBath}
	\Gamma^{\mathrm{ae}} = \kappa (2n^{\mathrm{th}} + 1),
\end{equation}
where $n^{\mathrm{th}}$ is the thermal population of a system part and $\kappa$ is its decay rate.
After many such processes, the wavefunction phases are no longer defined and the system can be described classically.

Besides, due to interactions in the model we consider, additional decoherence channels arise.
Due to the cross-Kerr coupling, fluctuations in one system gradually decohere another system.
Notably, that process rate grows with growing drive powers.
With a calculation similar to those of Refs.~\cite{wang2021quantum,gambetta2006qubit,clerk2010introduction,sokolov2023thesis}, we arrive at the estimate of
\begin{equation}
\label{eqDecoherenceCrossKerr}
	\Gamma^{\mathrm{cK}}_a = \frac{4 K^2}{\kappa_2} n_b (2n^{\mathrm{th}}_b + 1)
\end{equation}
for the associated decoherence rate of the mechanical system.
Mind the factor of two in the formula due to the fact that $\kappa_2$ denotes the FMR half-linewidth.
Here the magnon population $n_b$ due to the coherent drive is much larger than the thermal population $n^{\mathrm{th}}_b$.
The magnon number part of the expression comes from the mean-squared deviation $\mean{(b^\dag b - \mean{b^\dag b})^2}$ of a displaced thermal state with $b = b^{\mathrm{th}} + \sqrt{n_b}$.
A random change in the magnon number increases the phase uncertainty in the mechanical mode during the time $1/2\kappa_2$ when the magnon has not yet decayed.
Still, we consider the magnon coherence time to be small compared to the time of the cross-Kerr interaction $K^{-1}$.

Note that Eq.~\eqref{eqDecoherenceCrossKerr} overestimates the decoherence for more than a few thermal magnons~\cite{wang2021quantum,sokolov2023thesis};
a more intricate expression from Ref.~\cite{clerkutami2007using} should be used for quantitative estimates for higher temperatures.

In addition, there is also decoherence associated with the hybridization due to the linear interaction, similar to the Purcell decay~\cite{blais2004cavity,sete2014purcell,berman2011influence} in circuit QED.
In the non-resonant case its rate can be estimated as
\begin{equation}
\label{eqDecoherencePurcell}
	\Gamma^{\mathrm P}_a = \frac{2 g^2 \kappa_2}{(\omega_1 - \omega_2)^2}
							(2n^{\mathrm{th}}_b + 1)
\end{equation}
for the mechanical system.

Note that in the case multistability is pronounced in the system, irreversible quantum tunneling from a metastable well can be relevant.
That process only requires wavefunction coherence to persist on the short timescales of the inverse tunneling rate.
According to Ref.~\cite{andersen2020quantumvsclassical}, tunneling plays the role in the system relaxation if the cross-Kerr coupling is not much smaller than the decay rate, $K \gtrsim \kappa$, and the oscillator temperature is low enough.
We also expect that tunneling is dominated by the activation due to the shot noise from the coherent driving, if the drive is strong enough.

Generally, quantumness is a transient effect that usually fades away quite fast if no multistability is present.
In a given time, the classical limit can always be reached if the temperature is high or the oscillators are strongly driven.
Still, the respective populations should not be too high to invalidate our model Hamiltonian~\eqref{eqH}.
For the doubly-clamped magnetic beam, the amplitude of the magnetization precession should be much smaller than the saturation magnetization; and the beam oscillations should be small compared to its length (see Appendix~\ref{apMagnomechanicsDerivation} for details).

Finally, we provide numerical figures to assess the validity of the classical limit for the magnetic beam.
First of all, as we consider the case of fast damping of the FMR mode, multistability is not observable in our system.
Indeed, we expect the FMR half-linewidth of $\kappa_2 \approx 2\pi \, \qty 1{\GHz}$ for a beam made out of amorphous CoFeB~\cite{bilzer2006studycofeb}.
According to the estimates in Appendix~\ref{apMagnomechanicsDerivation}, the cross-Kerr coupling rate $K \sim 2\pi \, \qty{1}{nHz}$ for the beam cross-section of $\qty 1 {\micro\meter} \times \qty 5 {nm}$.
The cross-Kerr shift $Kn_a$ is responsible for the multistability, where $n_a$ is the vibrational mode occupancy.
The shift is negligible compared to the half-linewidth $\kappa_2$ when there is no more than $10^{17}$ phonons in the beam vibrational mode.
In that case there is no observable multistability, hence we can neglect tunneling for most practical cases.

Next, we observe that the decoherence due to the cross-Kerr interaction is completely negligible for such a small coupling and reasonable mode occupancy.
Indeed, according to Eq.~\eqref{eqDecoherenceCrossKerr}, the induced decoherence in the mechanical mode is of a rate $\Gamma^{\mathrm{cK}}_a / n_b \sim K^2 / \kappa_2 \sim 2\pi \, 10^{-27}\unit{\Hz}$ per one magnon from the coherent drive. 
Provided that the mechanical mode with frequency (see Appendix~\ref{apMagnomechanicsDerivation}) $\omega_a \sim 2\pi \, \qty 1 \MHz$ has~\cite{husain2003nanowire,kozinsky2006tuning,kozinsky2007nonlinear} $Q_a \sim 10^4$, it induces the FMR mode decoherence of a rate $\Gamma^{\mathrm{cK}}_b / n_a \sim Q_a K^2 / \omega_a \sim 2\pi \, 10^{-21}\unit{\Hz}$ per one coherent phonon.
Moreover, there is no decoherence due the Purcell decay as described by Eq.~\eqref{eqDecoherencePurcell}, as the beam symmetry precludes the linear interaction between the modes.

Therefore, if the coupling to the modes beyond our model is also small, the dominant mechanism for losing quantum coherence in the magnetic beam is the exchange of excitations with the environment.
According to Eq.~\eqref{eqDecoherenceBath}, at the room temperature $T$ the FMR mode quantum coherence vanishes after $1/\Gamma^{\mathrm{ae}}_b \sim \qty{0.1}{ps}$, where we have estimated the mode thermal occupancy $n_b^{\mathrm{th}} \approx k_B T / \hbar\omega_b \approx 600$ with $k_B$ the Boltzmann constant.
In the vibrational mode, coherence vanishes after $1/\Gamma^{\mathrm{ae}}_a \sim \qty 1{ns}$.
For such times, averaged mode observables do not exhibit any quantum effects. In other words, the classical limit where $\Gamma_{a/b}^{\mathrm{ae}} >  \omega_{a/b}$ is reached in the stationary state at sufficiently high temperatures.

In practice, quantum effects may vanish even on a shorter timescale as the above estimates do not include pure dephasing induced by the environment.

\section{Harmonic expansion of the equations of motion}
\label{secExpansion}

Here we obtain the equations for the harmonics of $a$ and $b$ by plugging the harmonic expansions of the system variables into the dynamic equations.
This approach is often called the harmonic balance method.
We assume that the frequencies of the modes of both system parts are non-equidistant, hence the higher modes are not excited by the harmonics.

The equation of motion for $b$ reads
\begin{multline}
\label{eqDiffb}
	\dot b = (-i\omega_2 - \kappa_2) b - 2if_2 e^{-i\Omega_2 t}
\\
			- ig(a + a^*) - {\frac 1 2} iK (b^* + b) (a^* + a)^2.
\end{multline}
Here $\kappa_2$ denotes the half-linewidth of the second oscillator.
The equation for $a$ has the same form, albeit with $\kappa_1$ denoting the half-linewidth of the first oscillator and similar changes.
Formally, the equations can be obtained for example with the Langevin approach.
That consists of adding the bath terms to the Hamiltonian~\eqref{eqH}, writing out the respective Langevin equations~\cite{scully1997quantum,berman2011influence,sokolov2020singlephoton} in the classical limit, and neglecting the thermal population of the modes.

In the steady state, the drives can only excite harmonics with frequencies
\begin{equation}
	\Omega_{kl} = k \Omega_1 + l \Omega_2
\end{equation}
with $k$ and $l$ integer.

We assume that the system does not undergo a subharmonic bifurcation~\cite{lukomsky2004cascades,chalyy2004nelinijni,burgelman2022subharmonicjj,alcorta2019perioddoubling} and subharmonics are negligible.
$a$ and $b$ can then be expanded in the form
\begin{equation}
\label{eqExpansions}
	a = \sum_{k,l = -\infty}^\infty A_{kl} e^{-i \Omega_{kl} t},
\quad
	b = \sum_{k,l = -\infty}^\infty B_{kl} e^{-i \Omega_{kl} t}
\end{equation}
in terms of the harmonic weights $A_{kl}$ and $B_{kl}$.
There is a symmetry in the harmonic weights $A_{kl} \leftrightarrow B_{lk}$ due to the symmetry $a \leftrightarrow b$, $\Omega_1 \leftrightarrow \Omega_2$, $f_1 \leftrightarrow f_2$, $\kappa_1 \leftrightarrow \kappa_2$, $\omega_1 \leftrightarrow \omega_2$ in the equations of motion: Eq.~\eqref{eqDiffb} for $b$ and the respective one for $a$.
As $a$ and $b$ contain higher harmonics, they generalize the concept of phasors known in the theory of linear electrical circuits;
however, we still call them phasors for simplicity.

We plug these expansions into the differential equation~\eqref{eqDiffb} for $a$ and the respective equation for $b$.
Then we collect the terms with the same exponents into algebraic equations.
For example, the equation on $B_{21}$ reads
\begin{dmath}
\label{eqB21_ordK}
(2\Omega_1 + \Omega_2 - \tilde\omega_2 + i\kappa_1) \, B_{21}
\\
	= g A_{\kl{-2}{-1}}^* + g A_{21}
	+ \frac 1 2 \, K\,B_{01}\,A_{10}^{2}+K\,B_{10}\,A_{01}\,A_{10}+K\,
 B_{\kl{-1}{0}}^*\,A_{01}\,A_{10}+K\,B_{10}\,A_{\kl{0}{-1}}^*\,A_{10}
 +K\,B_{\kl{-1}{0}}^*\,A_{\kl{0}{-1}}^*\,A_{10}

\end{dmath}
up to the terms that are linear in $K$.
Here $\tilde\omega_2$ denotes the resonance frequency as renormalized by the coupling;
we find an approximate expression for it in Sec.~\ref{secNonlinSdb}.
Consider the rhs of Eq.~\eqref{eqB21_ordK}.
The terms proportional to $g$ describe the linear processes that ``transfer'' the $A_{21}$ harmonic into the $B_{21}$ one.
The terms proportional to $K$ describe the cross-Kerr processes that generate the harmonic from the oscillations at the drive frequencies.
In all of these terms, the respective indices in all factors sum up to match those in the lhs, with the indices of a complex conjugate inverted.

We use Eqs.~\eqref{eqDiffb}--\eqref{eqExpansions} in Sec.~\ref{secNonlinSdb}.
In the next section, we consider a more general case than that given by Hamiltonian~\eqref{eqH} and the respective equation of motion~\eqref{eqDiffb}.
Still the expansions~\eqref{eqExpansions} are the same in that case.

\section{Coupling signatures in the harmonic structure}
\label{secNo11}
Here we discuss which harmonics are generated by a Kerr nonlinearity and compare to the harmonics generated by an optomechanical coupling.
We assume that subharmonics are not excited.

\subsection{No first sidebands with a cross- or self-Kerr nonlinearity only}
First we show that no combination of linear, Kerr, and cross-Kerr couplings generates a harmonic at the sum or difference of the driving frequencies $\Omega_1 \pm \Omega_2$.
The latter harmonics are often referred to as the (first) \emph{sidebands}.
In general, these types of coupling only generate harmonics at frequencies $k\Omega_1 + l\Omega_2$ with $k + l$ odd.

Any type of a Kerr coupling gives rise to the terms in the equations of motion for $a$ that are of the form
\begin{equation}
\label{eqKerrTerm}
	-i K c_{(*),(*),(*)} a^{(*)} b^{(*)} b^{(*)},
\end{equation}
where $c_{(*)(*)(*)}$ are constants that are different for each combination of complex-conjugated operators in a term.
They arise due to an expansion of a concrete form of the Kerr coupling.

\begin{figure}
\includegraphics{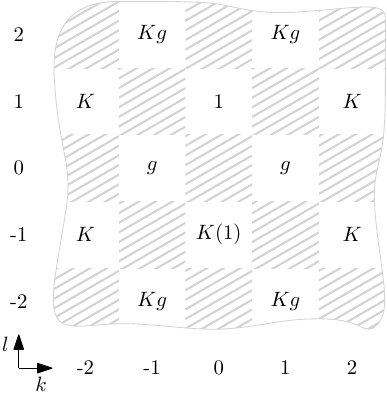}
\caption{Leading orders in the coupling constants of the first few harmonics $B_{kl}$ ($A_{lk}$) of $b$ ($a$) for the magnomechanical cross-Kerr coupling as in Hamiltonian~\eqref{eqH}.
The dashed spaces indicate absence of a harmonic;
their position is universal for any type of a Kerr coupling or any coupling of an odd power in the equations of motion.
In the white spaces the leading orders of a harmonic are indicated, in the coupling strengths $g$ and $K$.
The figure assumes the $\propto g^2$ contributions to be small compared to the $\propto K$ ones and vice versa.
The main contribution to $B_{\kl0{-1}}$ differs in the wide linewidth $\kappa_2 \gg K n_a$ limit close to the resonance $|\Omega_2 - \omega_2| \lesssim \kappa_2 $, where $n_a = a^* a$ is the mechanical mode population;
or further away from the resonance $|\Omega_2 - \omega_2| \gg \kappa_2, K n_a$.
The order relevant to these cases is indicated in the brackets.
Its calculation requires taking into account the co-rotating terms in the drive Hamiltonian.
}
\label{figCheckerboard}
\end{figure}

Let us check what harmonics are excited by the drives.
We consider the drives applied as described by Hamiltonian~\eqref{eqH} or any other case with two tones applied to the system.
The tones first excite the harmonics with only one ``1'' in the indices.
These are $A_{10}$ and $B_{01}$ for the driving as described by Hamiltonian~\eqref{eqH}.
A Kerr process generates a harmonic with indices that are a sum of indices of three other harmonics, according to Eq.~\eqref{eqKerrTerm} and the expansions~\eqref{eqExpansions}---see Eq.~\eqref{eqB21_ordK} for example.
As we start with the harmonics with an odd sum of indices, adding up three of them again gives an odd sum, and so on.
A linear process can only transfer a harmonic from one oscillator to another one preserving the harmonic indices [see the second term in the rhs of Eq.~\eqref{eqB21_ordK}] or inverting them~(the first term there).
None of that changes the sum parity.
Therefore, the drives only excite the harmonics with indices $k$ $l$ such that $k + l$ is odd.
The relevant harmonics form a checkerboard pattern in Fig.~\ref{figCheckerboard}.
Most notably, the first sidebands $A_{\kl1{\pm1}}$ and $B_{\kl1{\pm1}}$ are not excited.

Our considerations are also valid when there is a crosstalk between the drives.
In that case one may need to include the co-rotating part of the drive terms in Hamiltonian~\eqref{eqH}, as the conditions $\Omega_1 - \omega_2 \ll \Omega_1 + \omega_2$ and $\Omega_2 - \omega_1 \ll \Omega_2 + \omega_1$, that are analogous to conditions~\eqref{eqRWAdetuning}, can be broken.
While the crosstalk contributions to the populations are small compared to the resonant ones, they can become considerable in the nonlinear harmonics.
One possible mechanism for that is when the crosstalk contribution in $A_{10}$ becomes comparable to the non-resonant terms ``transferred'' from another mode $B_{01}$ by the linear coupling [see Eq.~\eqref{eqB21_ordK} for a similar process].
Harmonic terms of order $g$ or higher---e.g., in $B_{12}$ (see Fig.~\ref{figCheckerboard})---can be generated out of these transferred terms.
Therefore, the crosstalk also provides a sizeable contribution there.

The result of this section can be generalized to any coupling of odd power in the operators $a^{(\dag)}$ and $b^{(\dag)}$ in the equations of motion.
This is a generalization of a known fact that only odd harmonics are generated in a driven system with a nonlinearity of an odd power $x^{2n+1}$ in the system variable $x$~\cite{gandzha2004uniform}.
Note that if subharmonics become excited, they also excite the even harmonics.
This bifurcation was studied in Refs~\cite{lukomsky2004cascades,chalyy2004nelinijni} in a Duffing oscillator without damping.
While our system is related to it~\cite{manevitch2010limiting}, we expect that due to the damping subharmonics do not occur in the steady state of a moderately driven system~\cite{burgelman2022subharmonicjj}.
That assumption is even more likely to hold when in one of the modes damping dominates the nonlinearity.
We consider that case in Secs.~\ref{secNonlinSdb}--\ref{secAnticrossing}.

\subsection{Comparison with optomechanical coupling}

Now we compare the results above with the case when a nonlinearity of a previous order---the optomechanical coupling---is present.

Such a coupling arises in general cavity optomechanical systems where the electromagnetic field in the cavity interacts with the mechanical vibrations of one of the cavity mirrors~\cite{aspelmeyer2014optomechanics}.
It can also be encountered in the coupling between the mechanical vibrations and other dynamical modes, such as ferromagnetic resonance~\cite{zhang2016cavity,kansanen2021magnomechanics,potts2021dynamical,shen2022bistability}.
In a Hamiltonian, the respective interaction term is $\frac 1 2 G (b^* + b)^2 (a^* + a)$ with the optomechanical coupling strength $G$.
Its RWA part reads $G b^* b (a^* + a)$: in other words, the cavity or ferromagnetic resonance frequency is modulated by the amplitude of the mechanical oscillation.
The coupling gives rise to the term
\begin{equation}
	-i G (b^* + b)(a^* + a) 
\end{equation}
in the equation of motion for $b$.
By plugging the expansion~\eqref{eqExpansions} into that term, it can be seen that it generates harmonics which are either a sum or a difference of the input harmonics.
A plus or minus sign can occur in front of the sum or difference.
Linear sidebands $\Omega_1 \pm \Omega_2$ are thus excited directly by the drives, in the order $G$.
Moreover, all possible harmonics are excited, for example, by processes combining first sidebands with the drive tones again.
This can be seen as these processes making steps of $\pm 1$ in the harmonic indices.
In accordance with our reasoning, RWA analyses of Refs.~\cite{rabl2011photonblockade,nunnenkamp2011singlephoton,kong2018highorder} predict generation of all sidebands $\Omega_2 + n \Omega_1$ with $n$ integer.
Note that when $\Omega_1 = 2\Omega_2$ and one is only interested in the sideband at $\Omega_1 - \Omega_2 = \Omega_2$, which is at resonance, the RWA form of the coupling suffices.
In the case when the higher harmonics are of interest, the non-RWA form we consider is required.

\section{FMR sidebands in the cross-Kerr magnomechanics}
\label{secNonlinSdb}

As shown in the previous section, the cross-Kerr coupling does not generate the $\Omega_1 \pm \Omega_2$ sidebands.
The most prominent nonlinear harmonics are then of the order $K$ in the coupling constants~(see Fig.~\ref{figCheckerboard}).
We call these the second sidebands.
We focus on some of these harmonics in the magnomechanical system shown in Fig.~\ref{figBeam}.
As we show in detail in Appendix~\ref{apMagnomechanicsDerivation}, Hamiltonian~\eqref{eqH} applies in that case.
There, the mechanical mode of the magnetic beam~(with the dimensionless phasor $a$ in the Hamiltonian) interacts linearly and through the cross-Kerr coupling with the beam FMR mode (with the phasor $b$).
There is no coupling of the optomechanical form, because the symmetry of the mechanical mode precludes it.
The nonlinearity generates second FMR sidebands at $\Omega_2 \pm 2\Omega_1$ that can be picked up by an rf-antenna in the experiment.
In this section, we obtain quantitative analytical results for the magnitude of the second sideband in the limit of the wide FMR linewidth.

The second sideband magnitude can be measured in the picked up signal with a spectrum analyzer.
The magnitude of the blue second sideband is proportional to $|B_{21} + B_{\kl{-2}{-1}}^*|$ according to Eqs.~\eqref{eqExpansions}.
Equation~\eqref{eqB21_ordK} and a similar one for $B_{\kl{-2}{-1}}$ elucidate that the sideband magnitude is of order $K$, given that the drives directly excite $A_{10}$ and $B_{01}$.
In the limit of a wide FMR linewidth
\begin{align}
\label{eqFatFMR}
\nonumber
	\kappa_2 \gg {}&K n_a, \, n_a \sqrt{|K||\Omega_2 - \omega_2|}, \,
\\
				 {}&g^2 / |\Omega_2 - \omega_1|, \,
					g \sqrt{|\Omega_2 - \omega_2| / |\Omega_2 - \omega_1|},
\end{align}
the resonance shift due to the interaction is not resolved in the FMR.
Then in the leading order in $g$ and $K$, the second sideband generated at $ \Omega_2+2\Omega_1$ is
\begin{equation}
\label{eqB21_01}
	B_{21}\approx
 \frac {K\,A_{10}^{2}\,B_{01}} {2\,\left(2\,\Omega_1+\Omega_2-\omega_2+i\,\kappa_2\right)}
,
\end{equation}
where we neglect the terms of order $g^3$, $g^2 K$, and $K^2$.
In Eq.~\eqref{eqB21_01}, it is enough to take into account only the direct influence of the drives on the weights $A_{10}$ and $B_{01}$.
The latter can be found right away:
\begin{equation}
\label{eqB01}
	B_{01} \approx \frac{f_2}{\Omega_2 - \omega_2 + i\kappa_2}
\end{equation}
in the approximation of the wide FMR linewidth~\eqref{eqFatFMR}.
However, we assume that the linewidth of the mechanical mode is narrow enough to resolve the coupling-induced shifts in its resonance.
It is still possible to quickly obtain the correct result, that we later confirm with a meticulous calculation.
In the mechanical resonance shift, the resonant terms dominate;
therefore, in most of the cases one can drop the co-rotating terms in Hamiltonian~\eqref{eqH} in obtaining the shift.
Close to the linear resonances that are detuned from each other, this yields the renormalized frequency of the vibrational mode
\begin{equation}
\label{eqw1eff}
	\tilde\omega_1 \approx \omega_1+K\,\left|B_{01}\right|^2
,
\end{equation}
where we neglect the shift due to the linear coupling as it is proportional to $1/(\Omega_1 - \omega_2)$, which follows from the well-known expression for the normal mode of two linearly coupled oscillators.
Mechanical mode resonance shifts proportionally to the number of magnons due the cross-Kerr coupling.
Amplitude of the mechanical mode primary harmonic reads
\begin{equation}
\label{eqA10}
	A_{10} \approx \frac{f_1}{\Omega_1 - \tilde\omega_1 + i\kappa_1}.
\end{equation}
As $\tilde\omega_1$ shows up in the $A_{10}$ denominator, small changes in it influence the resonance shape that can be observed experimentally.
Thus it is interesting to obtain higher corrections to $\tilde\omega_1$.
Moreover, we can also determine the limits of applicability of the RWA we have used.

We proceed to obtain Eqs.~\eqref{eqw1eff}--\eqref{eqA10} with a more formal approach.
In the equation for $A_{10}$, we plug the expressions for $B_{kl}$ in such orders in $g$ and $K$ that the terms free of $A_{10}$ are determined in the orders $g^2$, $K^2$ and $gK$.
While doing that, we treat $B_{kl}$ as functions of $A_{mn}$ regardless of the order of magnitude of $A_{mn}$.
That way we obtain the renormalization in the $A_{10}$ resonance, including the terms nonlinear in $A_{10}^{(*)}$.
The respective coefficients of $A_{10}$, $A^*_{10}$ and any resulting combinations of them are determined up to and including the terms proportional to $g^2$, $K^2$ and $gK$.
We have performed that procedure for the harmonic weights with $|k|,|l| \leq 2$ in expansions~\eqref{eqExpansions} with a custom asymptotic solver~\footnote{The asymptotic solver codes can be accessed at \url{https://www.gitlab.com/matan-bebop/iterate-solve}.
It allows for solving quite a number of nonlinear equations (50 complex ones in the present work) with many terms (about 1000 for each equation) by performing symbolic iterations.
To perform the calculation within the limit of 16Gb of RAM available, we use the ``canonical'' representation~\cite{fateman1971macsyma} of sparse polynomials in both the solver and the application code.
The latter can be obtained by a personal inquiry.} in the \textsc{Maxima} computer algebra system.
Taking into account higher harmonics does not make sense in the second order in $g$ and $K$ we consider, as they are generated by at least three respective processes.
The resulting equation reads
\begin{widetext}
\begin{dmath}
\label{eqA10full}
\left(\Omega_1-\tilde\omega_1+i\,\kappa_1\right)\,A_{10} 
+\frac 1 2 \, A_{10}\,\left|A_{10}\right|^2\,\left|B_{01}\right|^2\,K^{2}\,\left(-\frac 1 {2\,\Omega_1-\Omega_2-\omega_2+i\,\kappa_2}-\frac 1 {2\,\Omega_1+\Omega_2-\omega_2+i\,\kappa_2}+\frac 2 {\Omega_2+\omega_2-i\,\kappa_2}+\frac 1 {2\,\Omega_1-\Omega_2+\omega_2+i\,\kappa_2}+\frac 2 {\Omega_2+\omega_2+i\,\kappa_2}+\frac 1 {2\,\Omega_1+\Omega_2+\omega_2+i\,\kappa_2}\right) 
\\ 
= f_1 
+\frac 1 2 KB_{01}^{2}\left(A_{\kl{1}{-2}}+A_{\kl{-1}{2}}^*\right) + K|B_{01}|^2\,A_{\kl{-1}{0}}^* + \frac 1 2 KB_{01}^{*2}\left(A_{12}+A_{\kl{-1}{-2}}^*\right)
,
\end{dmath}
\end{widetext}
with the effective frequency given by the full expression $\tilde\omega_1 = \omega_1+K\,\left|B_{01}\right|^2+\left(
 \frac 1 {\Omega_1-\omega_2+i\,\kappa_2}-
 \frac 1 {\Omega_1+\omega_2+i\,\kappa_2}\right)\,g^{2}
$.
The terms proportional to $g^2$ in that expression describe the frequency pull due to the linear interaction.
The first of these terms is similar to the dispersive shift in cavity quantum electrodynamics~\cite{blais2004cavity,sokolov2020singlephoton}.
It becomes comparable to the cross-Kerr shift in the case of a strong linear interaction, $g^2 \gtrsim K |B_{01}|^2 |\Omega_1 - \omega_2|$.
The second $\propto g^2$ term describes a Bloch-Siegert shift~\cite{forndiaz2010observation,sokolov2020singlephoton} in the FMR.
It becomes comparable to the dispersive shift when the mechanical drive is far detuned from the FMR, $|\Omega_1 - \omega_2| \sim \Omega_1 + \omega_2$.
We neglect the $\propto g^2$ terms in Eq.~\eqref{eqw1eff}, as we assume the mechanical drive at $\Omega_1$ detuned from the FMR at $\omega_2$, and $g$ sufficiently small.

Consider the other terms in Eq.~\eqref{eqA10full} that are proportional to $A_{10}$ and therefore explicitly renormalize the mechanical resonance.
These terms are given in the lhs of the equation.
The terms proportional to $K^2$ with the sum of frequencies in the denominator only matter for the ultrastrong cross-Kerr interaction, i.e., when $K|A_{01}|^2 {}\gtrsim \kappa_2, \, \omega_{1,2}, \, \Omega_{1,2}$.
Under the condition~\eqref{eqFatFMR} of wide FMR linewidth, this regime still can be reached if there are much more phonons than magnons $|A_{10}|^2 \gg |B_{01}|^2$.
The rest of the $\propto K^2$ terms can also become substantial close to the respective nonlinear resonances.
In any case, when these terms are non-negligible, multistability can arise, as they are nonlinear in the phonon number.

Now let us discuss the formally free terms that are given in the rhs of Eq.~\eqref{eqA10full}.
The first term there, $f_1$, describes the mechanical driving.
Usually it is the dominant term; we compare other terms to it.
The other terms stem from the cross-Kerr processes that contribute to $A_{10}$.
The term containing $A_{\kl{-1}0}$ becomes substantial away from the linear mechanical resonance.
The terms with the weights $A_{\pm 1 \pm 2}$ of the mechanical second sidebands should be accounted for when driving close to the respective nonlinear resonances.
When such a nonlinear resonance is close to the linear mechanical resonance, we expect them to repulse.
Indeed, one can check that the equations for $A_{\pm 1 \pm 2}$ depend on $A_{10}$.
That results in a repulsion of size $|K||B_{01}|^2$ similar to the avoided-crossing effect that we discuss in Sec.~\ref{secAnticrossing}.
Note that the terms of order $K^3$ that we neglect in the rhs provide a similar pull in the resonance frequency that can be of $K^2$ order.
However, we neglect this and similar shifts to the explicit $\sim K^2$ renormalization terms in the lhs.
These shifts are small because they arise due to the next-order weights $A_{mn}$,  and they are of higher order in the magnon population $|B_{01}|^2$.
Finally, neglecting all non-resonant terms in Eq.~\eqref{eqA10full} yields Eq.~\eqref{eqA10} that is valid far from higher resonances.

\begin{figure}
\includegraphics{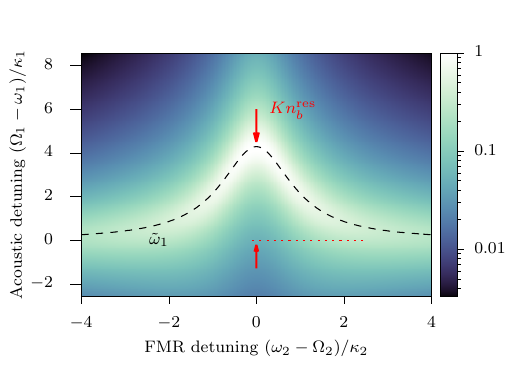}
\caption{Relative magnitude~\eqref{eqSecondBlueSideband} of the second blue sideband in the FMR, as a function of detunings.
In resonance with the FMR, the mechanical resonance $\tilde\omega_1$ seen in the sideband shifts proportionally to the cross-Kerr coupling strength $K$ and the resonant magnon population $n_b^{\text{res}}$.
Interaction-induced shifts of the FMR resonance are not resolvable under the condition~\eqref{eqFatFMR}.
The nonlinearities in Eq.~~\eqref{eqA10full} can be neglected, and the system is driven far from the nonlinear resonances.
}
\label{figSideband}
\end{figure}

Equations~\eqref{eqB21_01}, \eqref{eqB01}, and~\eqref{eqA10} constitute the explicit form of the $B_{21}$ expression.
$B_{\kl{-2}1}$, $B_{\kl2{-1}}$, and $B_{\kl{-2}{-1}}$ have an analogous form;
one can write them out by modifying the denominator accordingly and by choosing the conjugates in the rhs to match the harmonic indices in the lhs.
We also provide a general formal rule to obtain the weights with negative indices, which might be useful for more complicated expressions.
Equation~\eqref{eqDiffb} is invariant with respect to the substitution $b \to b^*$, $t \to -t$, $\Omega_2 \to -\Omega_2$, while $\kappa_2 \to -\kappa_2$.
Applying that to the $b$ expansion in Eqs.~\eqref{eqExpansions} and equating the result to the original expansion yields that
\begin{equation}
	B_{\kl{k}{-l}} = B^*_{kl}\big|_{\Omega_2 \to -\Omega_2,\, 
									\kappa_2 \to -\kappa_2}.
\end{equation}
An analogous rule holds for the first harmonic index.

One might wonder, why do we not take into account the co-rotating terms in the \emph{drive} Hamiltonian here?
Indeed, for a wide FMR linewidth, they should be accounted for to obtain the correct leading order of $B_{\kl0{-1}}$ and $A_{\kl{-1}0}$ (see Fig.~\ref{figCheckerboard}).
However, the drives contribute directly only to $B_{\kl0{-1}}$ and $A_{\kl{-1}0}$ through the co-rotating terms.
$B_{\kl0{-1}}$ and $A_{\kl{-1}0}$ then participate in processes that contribute to other harmonics---but their contribution is always negligible compared to that of $B^*_{01}$ and $A^*_{10}$ when the conditions~\eqref{eqRWAdetuning} hold.
The RWA drive Hamiltonian in Eq.~\eqref{eqH} works well for determining $B_{21}$ and other nonlinear harmonics.

In Fig.~\ref{figSideband}, we plot the magnitude
\begin{align}
\label{eqSecondBlueSideband}
	|B_{21} + B_{\kl{-2}{-1}}^*| &= |B_{21}|
		\Big|1 - \frac{2\Omega_1 + \Omega_2 - \omega_2 + i\kappa_2}
				      {2\Omega_1 + \Omega_2 + \omega_2 + i\kappa_2}\Big|
\\
\label{eqSecondBlueSidebandApprox}
	&\approx |B_{21}| \frac{\Omega_2}{\Omega_1 + \Omega_2}
\end{align}
of the second blue sideband in the FMR, using Eqs.~\eqref{eqB21_01}, \eqref{eqB01}, and~\eqref{eqA10} in the vicinity of the FMR and the mechanical resonance.
In Eq.~\eqref{eqSecondBlueSidebandApprox}, we neglect $i\kappa_2/2$ in comparison to $\Omega_1 + \Omega_2$.
According to the equations, the FMR line has a Lorentzian shape, and the mechanical resonance line is a squared Lorentzian.
Magnitude of the sideband is proportional to the mechanical driving power and the square root of the FMR driving power.
As demonstrated in the figure, the mechanical resonance shifts proportionally to $K$ and the number of magnons according to Eq.~\eqref{eqw1eff}.
When the Fig.~\ref{figSideband} plot is valid, the analogous plot for the magnitude of the second red sideband $|B_{\kl2{-1}} + B_{\kl{-2}1}^*|$ is visually indistinguishable from Fig.~\ref{figSideband};
although, its magnitude is approximately $(\Omega_1 + \Omega_2)/|\Omega_1 - \Omega_2|$ times smaller in the plot region.

There is a striking difference between our result and the first nonlinear harmonics predicted by using RWA in all terms of Hamiltonian~\eqref{eqH}.
In the all-RWA prediction, $B_{21}$ and $B_{\kl{-2}{-1}}$ vanish, up to and including the order of $g^2 K^2$.
Besides, $B_{\kl2{-1}}$ and $B_{\kl{-2}1}$ arise only in the $Kg^2$ order.
Moreover, the latter harmonics exhibit antiresonance at the resonant frequency of the mechanical mode: that is, a Lorentzian-shaped dip instead of a peak in Fig.~\ref{figSideband}.

Consider the assumptions that we use to plot Fig.~\ref{figSideband}.
First, the FMR nonlinear resonances $\omega_2 \approx 2\Omega_1 + \Omega_2, 3\Omega_1 + \Omega_2$ etc are avoided.
That is the case if $\Omega_2 \gg \kappa_1, |K||B_{01}|^2$.
That requirement also assures that the nonlinear frequency shifts in Eq.~\eqref{eqA10full} are off-resonant and can be neglected.
In the mentioned condition, we use that the span of the resonance features is about $|K||B_{01}|^2$ (see Fig.~\ref{figSideband} and Sec.~\ref{secAnticrossing}) or about its linewidth.
Analogously, we assume that the mechanical nonlinear resonances do not influence the plot.
Besides, we assume that the drives do not match.
We discuss the case of the matching drives in Sec.~\ref{secAnticrossing}.
In that case, the primary harmonic hybridizes with a second sideband, producing an avoided crossing.

Using the expressions for the first cross-Kerr sidebands, we next discuss a method to measure the cross-Kerr coupling strength $K$ or the FMR population $b^* b$.

\subsection{Possible measurement of the cross-Kerr coupling strength or magnon number: via the resonance shift in the FMR sideband}
\label{secMeasurement}

The magnitudes of the second sidebands are linearly proportional to the cross-Kerr coupling $K$, according to Eqs.~\eqref{eqSecondBlueSideband} and \eqref{eqB21_01} for the blue sideband and the similar ones for the red sideband.
However, to use that to determine $K$, one needs to know both the number of magnons and phonons, as well as the magnetization amplitude for a given magnitude of a signal picked up.
Here we describe how to estimate the cross-Kerr coupling $K$ using the shift in the mechanical resonance.
We assume that the FMR linewidth is wide and the condition~\eqref{eqFatFMR} holds.
We also assume that the linear FMR response is well-characterized, in particular that the number of magnons $n_b$ for a given drive strength is known.

One way to estimate $K$ is to inspect an FMR second sideband in the vicinity of the FMR and the mechanical mode resonances.
Figure~\ref{figSideband} shows that the mechanical resonance $\tilde\omega_1$ shifts linearly with the number of magnons $n_b = B_{01}^* B_{01}$ proportionally to $K$, according to Eq.~\eqref{eqw1eff}.
The shift $\tilde\omega_1 - \omega_1$ is largest at the FMR resonance $\Omega_2 = \omega_2$.
To estimate $K$, one measures the shift with respect to the FMR drive power calibrated in terms of the magnon number.
Then $K$ can be obtained from the slope of the shift dependence.
Vice versa, if $K$ is known, one can perform a similar measurement to measure the magnon number.

Alternatively, one can measure the shift directly in the resonant harmonic $A_{10}$~\eqref{eqA10} of the mechanical mode.
Similar measurements were performed in circuit QED in the dispersive limit to count photons in a microwave resonator~\cite{schuster2005ac,johnson2010qnd} and to readout a qubit~\cite{blais2004cavity,walter2017rapid,heinsoo2018rapid}.
However, in the magnomechanical system such a straightforward measurement might be harder to perform.
The shift is visible only in the narrow-linewidth mechanical mode.
Measuring it might add an unwanted complexity to an experimental setup.
One needs to drive both modes and pick up the mechanical mode displacement.
The latter requires conversion to an electric signal.
The acoustic drive port can be used for that, but then a circulator is needed to separate the reflection from the drive signal.
Adding another port for readout might complicate the setup as well.

In the next section, we discuss another possibility to measure the cross-Kerr renormalization $K n_b$ by the spectroscopic methods.

\section{Cross-Kerr avoided crossing between the linear mechanical resonance and the second-sideband resonance in hybrid systems}
\label{secAnticrossing}

We have already commented that Eq.~\eqref{eqA10full} for the primary harmonic $A_{10}$ depends on $A_{\kl{-1}2}$ and $A_{\kl1{-2}}$ which determine the second sideband of the mechanical mode.
Here we study the avoided crossing due to that coupling between harmonics.
We also argue that the interaction between $A_{10}$ and $A_{\kl{-1}2}$ is the most relevant for observations.
As in the previous section, we assume that the resonance of one of the two coupled systems is wide, according to Eq.~\eqref{eqFatFMR}.

Consider the mechanical mode.
Besides the linear resonance $\Omega_1 = \tilde\omega_1$ in its primary harmonic $A_{10}$, there are nonlinear resonances that occur when the higher harmonics attain their local maxima.
In the FMR mode, such a nonlinear resonance at $\omega_2 = \Omega_2 + 2\Omega_1$ produces its second sideband~\eqref{eqB21_01}.
Analogously, one of the mechanical mode second sidebands arises at $\tilde\omega_1 = -\Omega_1 + 2\Omega_2$.
This nonlinear resonance matches the linear one when $\Omega_1 = \Omega_2$.
We study the system response near that point.

We obtain the equations for $A_{10}$ and $A_{\kl{-1}2}$ similarly to Eq.~\eqref{eqA10full}.
They read
\begin{gather}
\label{eqA10anticrossing}
	\delta_1 A_{10} = f_1 + \varkappa_{11} A^*_{\kl{-1}2},
\\
\label{eqAm12anticrossing}
	\delta_{\kl{-1}2} A_{\kl{-1}2} = \varkappa_{11} A^*_{10},
\end{gather}
where we have defined
\begin{gather}
\label{eqd1dm12}
	\delta_1 = \Omega_1 - \tilde\omega_1 + i\kappa_1,
\quad
	\delta_{\kl{-1}2} = -\Omega_1 + 2\Omega_2 - \tilde\omega_1 + i\kappa_1,
\\
	\varkappa_{11} = K B_{01}^2 / 2.
\end{gather}
We have neglected the non-resonant terms, as well as the terms $-g^2 / (\Omega_1 - \omega_2 + i\kappa_2)$ and $-K^2 |A_{10}|^2 |B_{01}|^2 / (2\Omega_1 - \Omega_2 - \omega_2 + i\kappa_2)$ that appear in both $\delta_1$ and $\delta_{\kl{-1}2}$.
The latter become significant closer to the anticrossing point at $\Omega_1 = \Omega_2 = \omega_1 = \omega_2$.
The solution of Eqs.~\eqref{eqA10anticrossing}--\eqref{eqAm12anticrossing} is
\begin{equation}
\label{eqA10SolutionAnticrossing}
	A_{10} = \frac{f_1 \delta^*_{\kl{-1}2}}
				  {\delta_1 \delta^*_{\kl{-1}2} - |\varkappa_{11}|^2},
\quad
	A_{\kl{-1}2} = \frac{-f_1 \varkappa_{11} }
				  		 {\delta_1 \delta^*_{\kl{-1}2} - |\varkappa_{11}|^2}
\end{equation}
except for the anticrossing point $\Omega_1 = \Omega_2$.
Harmonic weights $A_{\kl{-1}0}$ and $A_{\kl1{-2}}$ can be obtained by using the index inversion rule provided in Sec.~\ref{secNonlinSdb}.

\begin{figure}
\includegraphics{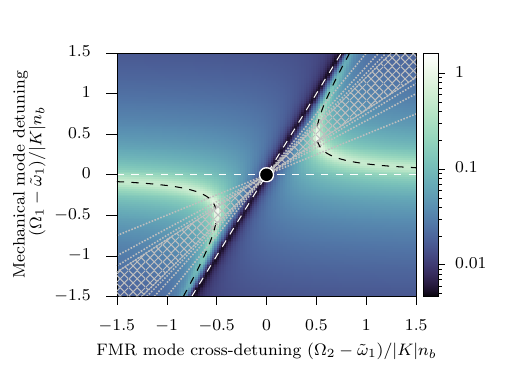}
\caption{Avoided crossing between the linear mechanical resonance and the second-sideband nonlinear resonance.
Color indicates the magnitude of the primary harmonic $|A_{10} + A^*_{\kl{-1}0}|$ [see Eqs.~\eqref{eqA10SolutionAnticrossing}] normalized by its approximate maximal value, $1/\kappa_1$.
It is plotted as a function of detunings from the mechanical resonance.
Resonance positions for small dissipation (black dashed) approach linear asymptotes (white dashed) far from the anticrossing.
Our theory breaks close to the anticrossing point (circle).
Higher resonances with the asymptotes (gray dotted) can be observable in experiment.
Their density increases (crossed area) towards infinity near $\Omega_1 = \Omega_2$.
Here the mechanical mode half-linewidth is $\kappa_1 = 0.05 \, |K|n_b$, where $n_b$ is the number of magnons.
Mechanical resonance frequency is $\tilde\omega_1 = 10^6 \, |K|n_b$.
Detuning from the FMR is not very large according to Eq.~\eqref{eqRWAdetuning}.
Interaction-induced shifts of the FMR resonance are not resolvable under the condition~\eqref{eqFatFMR}.}
\label{fig11Anticrossing}
\end{figure}

Plot of the mechanical mode primary harmonic magnitude $|A_{10} + A_{\kl{-1}0}|$ in Fig.~\ref{fig11Anticrossing} exhibits an avoided crossing.
It occurs due to the cross-Kerr coupling that results in the resonances shift, similarly to the case of an avoided crossing due to the linear interaction.
There is an antiresonance in the primary harmonic---the minimum in its magnitude---where all of the drive energy ends up in the sideband.
Plot of the second sideband magnitude $|A_{\kl{-1}2}+A_{\kl1{-2}}|$ looks similar, except for the lack of an antiresonance.
Positions of the plot features do not depend on the $\tilde\omega_1 / |K| n_b$ ratio of the mechanical resonance frequency to the cross-Kerr shift.
In the figure, we also show the position of the resonances in the low-dissipation limit when $\kappa_1^2 \ll |\varkappa_{11}|^2$.
In that case, they take place when
\begin{equation}
	\Omega_1 
		\approx \Omega_2 
				\pm \sqrt{(\Omega_2 - \tilde\omega_1)^2 - K^2 n_b^2 / 4}.
\end{equation}
Two resonance branches are separated by $Kn_b$.
Far from the anticrossing point the harmonics do not hybridize substantially, and the resonances approach $\Omega_1 = \tilde\omega_1$ and $\Omega_1 = 2\Omega_2 - \tilde\omega_1$.

In Eqs.~{\eqref{eqA10anticrossing}--\eqref{eqAm12anticrossing}, we have neglected the $-g^2 / (\Omega_1 - \omega_2 + i\kappa_2)$ and $-K^2 n_a n_b / (2\Omega_1 - \Omega_2 - \omega_2 + i\kappa_2)$ terms in the frequencies.
Far from the internal 1:1 resonance when $\omega_1 \approx \omega_2$, these terms are non-resonant.
Consider also the mentioned case of internal resonance.
Let us check whether our predictions of resonance repulsion in Eqs.~\eqref{eqA10anticrossing}--\eqref{eqAm12anticrossing} and Fig.~\ref{fig11Anticrossing} are not influenced by these terms.
The terms provide both real and imaginary corrections to the frequencies.
At the distance of about $|K|n_b/2$ from the anticrossing point, the real $\propto K^2$ correction is bounded as $|\mathrm{Re}(\propto K^2) / K n_b| < |K|n_a/2\kappa_2$.
Thus it is negligible compared to the shift $Kn_b$ in the limit of wide FMR linewidth~\eqref{eqFatFMR}.
The $\propto g^2$ real correction is negligible if $4g^2 \ll K^2 n_b + \kappa_2^2$.
The imaginary parts of the correcting terms renormalize the mechanical mode damping.
They can be approximated as $\mathrm{Im}(\propto K^2) \approx \kappa_2 \frac{n_b}{n_a} / \big(\frac{n_b^2}{4n_a^2} + \frac{\kappa_2^2}{K^2 n_a^2}\big)$ and $\mathrm{Im}(\propto g^2) \approx g^2 \kappa_2 / (\kappa_2^2 + K^2 n_b^2 / 4) < g^2 / |K|n_b$.
The first one is negligible compared to $Kn_b$ when $n_b \ll 2\kappa_2 / |K|$ and the second one is negligible when $g^2 \ll K^2 n_b$.

We have ignored the resonances in the sidebands beside the second ones.
We identify other resonance lines that cross the $\Omega_1 = \Omega_2 = \tilde\omega_1 = \omega_2$ point if the interaction between them is disregarded.
They are $\Omega_1 = [(k+1)\Omega_2 - \tilde\omega_1] / k$ and $\Omega_1 = [(k-1)\Omega_2 + \tilde\omega_1] / k$ for $k = 2, 3, \ldots$
We show these lines in Fig.~\ref{fig11Anticrossing}.
The nonlinear resonances occur in the harmonic weights $A_{\kl{-2}3}, A_{\kl{-3}2}, \ldots$ and $A_{\kl2{-1}}, A_{\kl3{-2}}, \ldots$
The weight $A_{\kl2{-1}}$ interacts with $A_{10}$ and $A_{\kl{-1}2}$ beyond the order in $g$ and $K$ that we consider in Eqs.~\eqref{eqA10full}, \eqref{eqA10anticrossing}--\eqref{eqAm12anticrossing}.
Hence it does not substantially influence the positions of the features in Fig.~\ref{fig11Anticrossing}.
Moreover, as $A_{\kl2{-1}} \sim K^2$ while $A_{10} \sim 1$, the respective resonance line should be relatively faint.
However, it is on par with other resonances in the plot of the second sideband magnitude $|A_{\kl{-1}2} + A_{\kl1{-2}}|$ similar to Fig.~\ref{fig11Anticrossing}.
All higher harmonic weights are of increasingly higher order in the coupling rates; the same holds for their interactions with $A_{10}$ and $A_{\kl{-1}2}$.
However, when approaching $\Omega_1 = \Omega_2$, the density of the higher resonances tends to infinity.
Calculating the combined effect of many nonlinear resonances in that region is beyond the scope of this work.
Still, further away from there we expect only minor modifications of the picture in Fig.~\ref{fig11Anticrossing}.

Other interactions involving only the primary and the second harmonics require rather exotic conditions to observe them.
We have identified the interacting harmonics with the equations similar to Eq.~\eqref{eqA10full}.
We obtain them for each harmonic with computer algebra as before.
Some of the interactions, like $A_{\kl{-1}{-2}}$ coupling to $A_{\kl{-1}0}$, involve the harmonics that are always off-resonant and thus hard to observe.
To observe $A_{\kl{-1}0}$ coupling to $A_{10}$, either the ultrastrong cross-Kerr coupling is required with $\tilde\omega_1 \lesssim |K|n_b$, or the mode should be overdamped with $\tilde\omega_1 \lesssim \kappa_1$, or its driving should be extremely off-resonant with $\Omega_1 \lesssim \max\{|K|n_b, \kappa_1\}$.
Other interactions, like $A_{\kl{-1}2}$ and $A_{\kl1{-2}}$, require similar conditions for the FMR frequencies.
Moreover, under the same condition, several interacting pairs are excited.
That can complicate the analysis of the observed data even if the required condition is met.

\section{Conclusion and outlook}
\label{secConclusion}

We have shown that for any type of self- or cross-Kerr coupling being the dominating nonlinearity---that is, when optomechanical coupling is negligible---first sidebands are absent in the system of two coupled oscillators driven by two tones.
In fact, only a specific set of harmonics is generated, as shown in Fig.~\ref{figCheckerboard}.
This is in contrast to optomechanical coupling that generates all possible harmonics, including the first sidebands.
While the first sidebands do appear if the system undergoes a subharmonic bifurcation~\cite{lukomsky2004cascades,chalyy2004nelinijni}, we expect that it does not occur in the steady state of a moderately driven system.
Hence, the absence of the first harmonic indicates that there is no optomechanical coupling in the system.
This result holds for nonlinear systems of any nature.

Then we have considered a magnomechanical system, where a mechanical mode couples to an FMR mode linearly and with a cross-Kerr coupling.
Optomechanical coupling is negligible in that system.
Our theory provides quantitative predictions of the second sideband magnitudes.
That requires taking the non-RWA coupling terms into account.
Our analytical results are valid for the usual case of where the FMR modes have a relatively large linewidth. 
Using our theory, we have discussed how to deduce the cross-Kerr coupling strength.
That can be achieved by measuring the dependence of the resonance shift in an FMR second sideband on the magnon population.
We also predict a nonlinear anticrossing between the primary and the higher resonance.
The cross-Kerr coupling strength or the magnon population can also be inferred from the gap between the resonance branches there.

We have shown in Appendix~\ref{apMagnomechanicsDerivation} that a doubly clamped magnetic beam in a symmetrical vibration mode is described with a cross-Kerr Hamiltonian.
We provide a quantum Hamiltonian, albeit it has been obtained with the semiclassical phenomenological approach.
In the main part of the paper, we work in the classical limit.
For the beam, it is a good approximation for high temperatures and when the losses in the FMR mode preclude multistability.
According to our derivation, linear coupling is absent in the symmetrical doubly clamped magnetic beam.
Still, we include it in the Hamiltonian in the main part of the paper for generality.
As the cross-Kerr coupling strength scales inversely to the beam width and the square of its height, smaller bridges are favorable for achieving a higher coupling rate.
We have estimated in Appendix~\ref{secEstimateK}, that the cross-Kerr coupling strength is at least in the nanohertz range for a CoFeB bridge with $\qty1{nm} \times \qty5{\micro\meter}$ cross section and negligible initial tension.
Surprisingly, in the bridges with higher tensions, the coupling strength can be up to four orders of magnitude higher.
It is promising to study whether a much higher coupling rate is achievable for a beam made out of a 2D magnetic material, such as~\cite{huang2017layer2d} monolayer $\mathrm{CrI_3}$, or $\mathrm{CrCl_3}$ and the related compounds~\cite{tartaglia2020accessing2d}.

We also mention side results of this work.
We have identified the renormalization of the beam tension due to its static magnetization.
As the frequency of the beam vibrations depends on the tension, one can tune it \emph{in situ} by rotating the magnetization with an external magnetic field.
Also, it is known that a large enough negative tension buckles the beam~\cite{kansanen2021magnomechanics}.
According to Eqs.~\eqref{eqNoBuckling} and~\eqref{eqMagneticTension}, an initially unbuckled beam close to the buckling transition can buckle when a magnetic field is applied.
For the materials with positive magnetoelastic constant or when the magnetoelasic energy is much smaller than the demagnetizing energy, the field that is transverse to the beam plane matters.
In the opposite case, the field along the beam can also buckle it.
The buckling can be larger than 1 nm for a CoFeB beam~\cite{kansanen2021magnomechanics}.
Then the magnetoelastic and the demagnetizing energies overcome the exchange energy, and the beam magnetization becomes inhomogeneous as discussed in Appendix~\ref{secHomogeneousM}.
It can be interesting to study the buckling of a magnetic beam as an orientational phase transition~\cite{landau1984electrodynamics}, similarly to Ref.~\cite{kalita2023spontaneous}.

It is also intriguing to use a nonlinear theory similar to ours for applications in the superconducting quantum computing.
Second sidebands might be used to characterize the cross-Kerr coupling in the computational basis of transmon qubits.
For that, one can measure the transmon response to continuous coherent driving.
The measurement should be able to detect second sidebands in the response.
We expect that only minor modifications to our theory are required to describe that case.
Indeed, a transmon in the computational basis is, to a good approximation, a resonator with a self-Kerr nonlinearity~\cite{rasmussen2021sccompanion,koch2007transmon}.
Besides adding this nonlinearity to Hamiltonian~\eqref{eqH} and the theory of Sec.~\ref{secNonlinSdb}, one should carefully treat the non-RWA nonlinear terms that are usually~\cite{ciani2019hamiltoncompcpbcrosskerr,rasmussen2021sccompanion} neglected.
As we have discussed, these terms are important for determining the nonlinear response.
Also, for the strong damping approximation of Sec.~\ref{secNonlinSdb} to hold, one would need to artificially decrease the relaxation time $T_1$ of one of the transmons; that can be done, for example, by coupling that transmon to a resonator so that the associated Purcell decay~\cite{sete2014purcell} is strong enough.
A cross-Kerr--type interaction also naturally arises due to the inductive coupling of flux qubits~\cite{orlando1999threejjqubit,rasmussen2021sccompanion,wendin2007qubits};
however, due to their high anharmonicity, this interaction is effectively linear.

It might be interesting to obtain our results applying approximate canonical transformations on the Hamiltonian~\cite{sedletsky2022hamiltoniankg,craig2010hamiltoniansw}.
Comparing to the approach of the present paper, we expect that using Hamiltonian transformations is much less computationally expensive.
Moreover, this approach can be easily extended to a quantized Hamiltonian using unitary transformations.

\begin{acknowledgments}
We thank Mika Sillanpää and Mishra Harshad for stimulating discussions.
Andrii Sokolov thanks Richard Fateman, Stavros Macrakis, and others from the \textsc{maxima-discuss} mailing list for help with various aspects of using the \textsc{Maxima} computer algebra system.
Andrii Sokolov also thanks Ivan Gandzha for enlightening comments on the subharmonic bifurcation.
This work was supported by the Academy of Finland (Contract No.~321982).
\end{acknowledgments}

\appendix

\section{Derivation of the nonlinear coupling terms for the doubly clamped magnetic beam}
\label{apMagnomechanicsDerivation}

Here we derive the quantum Hamiltonian of the doubly clamped magnetic beam.
We consider its first bending mode interacting with the homogeneous FMR mode.
Our treatment follows Ref.~\cite{kansanen2021magnomechanics}.
However, unlike that work that concentrates on a buckled beam, we study the case of initially unbuckled beam that vibrates symmetrically.
We show that a cross-Kerr coupling arises between the FMR and the mechanical modes;
we estimate the coupling strength.
Besides, we identify the influence of the beam magnetization on its tension and on the buckling transition.

\subsection{Beam vibrations}

First we consider the beam vibrational mode around its static configuration.
Here we only take into account the influence of the static magnetization.
We take into account the interaction with the dynamical part of the magnetization when considering the FMR mode.

\begin{figure}[b]
\includegraphics{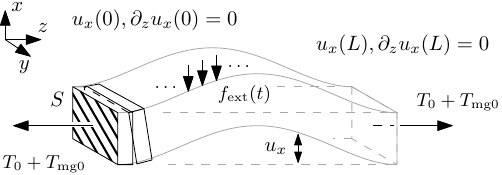}
\caption{The beam with a cross section area $S$ bends (solid gray) from its static configuration (dashed gray).
Each its point at a position $z$ along the beam is displaced by $u_x(z)$.
Elastic energy arises due to the ``overlap'' and ``stretching'' of infinitesimal beam cross-sections (solid black) with respect to each other.
Boundary conditions indicated in the figure describe the beam clamped at its ends.
An initial tension $T_0$ acts on the beam together with the tension $T_{\mg0}$ due to the static magnetization.
An external time-dependent force $f_{\ext}$ drives the beam vibrations.}
\label{figMechanics}
\end{figure}

We assume that the beam deflects only in the direction perpendicular to its equilibrium plane ($yz$-plane).
Beam displacements $u_x$ along the $x$-axis~(see Fig.~\ref{figMechanics}) are described by the equation~\cite{landau1984elasticity}
\begin{equation}
\label{eqBeamVibrations}
	\rho S \ddot u_x + E I_y \partial_z^4 u_x
		= (T_0 + T_{\mg0}) \partial_z^2 u_x + f_{\ext}(z,t).
\end{equation}
We have assumed that the deformations of the beam are small and $\partial_z u_x, L \partial^2_z u_x \ll 1$.
As already mentioned in Fig.~\ref{figBeam}, the beam is considered to be thin.
Here $S$ is the beam cross-section area, $\rho$ is its mass density, $EI_y$ is its bending stiffness with $E$ the Young's modulus and $I_y = wh^3/12 $ the area moment of inertia.
Note that $I_y$ here is of elastic origin and coincides with the moment of inertia of the thin plate only because of the formal similarities in the theories.
$f_{\ext}$ is a time-dependent force acting on the beam per unit length.
The force can be the radiation pressure force in a laser setup.
Alternatively, one can arrange an ac current to flow along the beam.
In the static magnetic field, a Lorentz force then arises acting on the beam normally to its plane.
While its direction changes when the beam vibrates, for small deformations it can be described by the same driving term in the Hamiltonian~\cite{kansanen2021magnomechanics}.
The initial tension $T_0$ is known to arise in doubly clamped beams~\cite{husain2003nanowire,kozinsky2006tuning,kozinsky2007nonlinear}.
Additionally, the tension $T_{\mg0}$ of magnetic origin should be taken into account.
To see that clearly, we write out the Hamiltonian that corresponds to Eq.~\eqref{eqBeamVibrations}:
\begin{widetext}
\begin{equation}
\label{eqHMechanical}
	H_{\vib} = \int_0^L dz \Big[\frac 1 2 \frac{\pi_x^2}{\rho S}
							- \frac 1 2 E I_y (\partial_z^2 u_x)^2
							+ \frac 1 2 T_0 (\partial_z u_x)^2
							- f_{\text{ext}} u_x \Big]
			   + F_{\dm0} + F_{\me0}
\end{equation}
\end{widetext}
with the canonical momentum $\pi_x = \rho S \dot u_x$.
Below we show that the static parts of the magnetic free energy satisfy $F_{\dm0} + F_{\me0} = \frac 1 2 T_{\mg0} (\partial_z u_x)^2$.
Comparing that to the third term in the integral in Eq.~\eqref{eqHMechanical}, we conclude that they provide an additional tension $T_{\mg0}$ in the beam.

Let us comment on the different terms of Eqs.~\eqref{eqBeamVibrations}--\eqref{eqHMechanical}.
The first term in the integral in Hamiltonian~\eqref{eqHMechanical} is the kinetic energy of a thin cross-section of the beam.
The second term there is the elastic energy density due to the beam curving (see Fig.~\ref{figMechanics} for interpretation).
When the tension is high or the beam stiffness $EI_y$ is small, that term is negligible to the tension terms.
Equation~\eqref{eqBeamVibrations} then becomes the usual wave equation of a vibrating string.

We assume that the beam is initially unbuckled.
For that, the beam should not be ``squeezed'' too much if the sum of initial tension and the tension due to the static magnetization is negative~\cite{kansanen2021magnomechanics}:
\begin{equation}
\label{eqNoBuckling}
	\frac {L^2}{EI_y} (T_0 + T_{\mg0}) > -4\pi^2.
\end{equation}

One can expand the beam vibrations into the normal modes with
\begin{equation}
\label{eqVibModesExpansion}
	u_x(z) = \sum_k x_k \upsilon_k(z),
\quad
	\int_0^L dz \upsilon_k \upsilon_l = L\delta_{kl}.
\end{equation}
Hamiltonian~\eqref{eqHMechanical} with no drive ($f_{\ext} = 0$) then becomes~\cite{kansanen2021magnomechanics} $H_{\vib0} = \sum_k ( \frac{p_k^2}{2m} + \frac 1 2 m k^2 \omega_k^2 x_k^2)$, where $m = \rho LS$ is the beam mass, $\omega_k$ is the resonance frequency of the $k$-th mode, and $p_k = m\dot x_k$ is the respective momentum.
We assume that only the first bending mode of the beam is excited.
Then the Hamiltonian of its free vibrations is simply $H_{\vib0} \approx \frac{p^2}{2m} + \frac 1 2 m \omega_\vib^2 x^2.$
The drive part of Hamiltonian~\eqref{eqHMechanical} becomes $H_{\vib\dr} = -x\int_0^L dz \, \upsilon f_{\ext}(t)$.
Here and below $\upsilon(z)$ denotes the normalized shape of the first bending mode from expansion~\eqref{eqVibModesExpansion}.

One readily quantizes the vibrational mode.
We promote the canonical coordinates $x$ and $p$ to operators with the commutation relation $[x, p] = i\hbar$.
It is convenient to introduce the phonon creation $a^\dag$ and annihilation $a$ operators with
\begin{gather}
\label{eqXinTermsOfa}
	x = x_\ZPF (a + a^\dag),
\quad
	p = \frac{-i\hbar}{2x_\ZPF} (a^\dag - a),
\\
	x_\ZPF = \sqrt{\frac{\hbar}{2m\omega_\vib}}.
\end{gather}
Here $x_\ZPF$ denotes the magnitude of zero-point fluctuations~\cite{clerk2010introduction} in the beam deflection $x$.
In terms of these operators, the Hamiltonian of the free vibration mode reads
\begin{equation}
\label{eqHFreeVibrationsSecondQuantized}
	H_{\vib0} = \hbar \omega_\vib a^\dag a.
\end{equation}
The drive Hamiltonian reads
\begin{equation}
\label{eqHdVib}
	H_{\vib\dr} = -\hbar x_{\ZPF} (a^\dag + a)
					 \int_0^L dz \, \upsilon f_{\ext}(t).
\end{equation}

Note that we have quantized the normal modes of  Eq.~\eqref{eqBeamVibrations} while taking into account the tension-dependent terms.
If we leave them out from the normal modes and quantize, squeezing~\cite{scully1997quantum} terms of the form $\sum_{kl} s_{kl} (a_k + a_k^\dag) (a_l + a_l^\dag)$ arise in the Hamiltonian, where we denote each mode phonon annihilation operator by $a_k$.
However, in the limit that recovers the string equation, the tension term is the only one to be accounted for in the potential energy in Hamiltonian~\eqref{eqHMechanical}.
In that case, it should be included in the normal mode calculation and enters the respective quantum operators.
We therefore conclude that the same should be done in the general case when both the tension and the elastic energies matter.

\subsection{Hamiltonian formalism for the magnetization precession}
\label{secHamiltonianStructure}

In the phenomenological approach, magnetization dynamics is governed by the Landau-Lifshitz-Gilbert equation~\cite{landau1980statisticalii}
\begin{equation}
	\dot{\vec M} = \mu_0 \gamma [\vec H_{\eff} \times \vec M]
				   + \frac\alpha{\Ms} [\vec M \times \dot{\vec M}],
\end{equation}
where $\mu_0$ is the magnetic permeability of the vacuum, $\gamma > 0$ is the gyromagnetic ratio of the medium, $\alpha > 0$ is its Gilbert constant that describes damping, and $\Ms = |\vec M|$ is the saturation magnetization of the magnet.
At each instant of time, magnetization $\vec M$ precesses around the direction of the effective magnetizing field
\begin{equation}
\label{eqLLG}
	 \vec H_{\eff} = - \frac 1 {\mu_0 V} \frac {\partial F}{\partial \vec M}
\end{equation}
which is determined from the medium free energy $F$.
Here we assume that the magnetization is homogeneous.
We evaluate this assumption in Sec.~\ref{secHomogeneousM}.

We consider small-angle precession around the static magnetization direction.
In that case damping does not depend on the precession angle.
We set the damping term in Eq.~\eqref{eqLLG} to zero and re-introduce the damping in the equation of motion~\eqref{eqDiffb}.
Now Eq.~\eqref{eqLLG} has a simple corresponding Hamiltonian.
With no loss of generality so far, we assume that the static magnetization is aligned along the $z$-axis.
We use that
\begin{equation}
\label{eqVectorApproximation}
	M_z^2 \approx \Ms^2 - M_x^2 - M_y^2,
\end{equation}
and the free energy becomes a function of $M_x$ and $M_y$ only.
Approximating $M_z \approx \Ms$ in the rhs of the Landau-Lifshitz equation~\eqref{eqLLG}, it can be rewritten as
\begin{align}
	\frac d{dt} (\zeta M_x)
		= {}& \{ \zeta M_x, F \}
		= - \frac{\partial F}{\partial (\zeta M_y)},
\\
	\frac d{dt} (\zeta M_y)
		= {}& \{ \zeta M_y, F \}
		= \frac{\partial F}{\partial (\zeta M_x)},
\end{align}
where $\zeta = \sqrt{V/{\gamma\Ms}}$ and $\{f,g\} = \frac{\partial f}{\partial (\zeta M_y)} \frac{\partial g}{\partial (\zeta M_x)} - \frac{\partial f}{\partial (\zeta M_x)} \frac{\partial g}{\partial (\zeta M_y)}$ denotes the Poisson bracket with respect to the canonical coordinate $\zeta M_y$ and the canonical momentum $\zeta M_x$.
Free energy $F$ plays the role of the Hamiltonian in the equations.

Now we can perform the canonical quantization.
$M_x$ and $M_y$ are promoted to operators that satisfy the commutation relation
\begin{equation}
\label{eqMyMxCommutator}
	[M_y, M_x] \approx i \hbar \gamma\Ms/V.
\end{equation}
The latter is determined from the usual commutator for the canonically conjugated variables.
Relation~\eqref{eqMyMxCommutator} is an approximation to the exact commutator of spin operators with the $z$-component of magnetization in the rhs~\cite{landau1980statisticalii}.
Note that, unlike the treatment of the mechanical mode, we do not include the terms similar to the tension terms in Eq.~\eqref{eqHMechanical} into the modes that are quantized.
As we discuss later, these terms provide quantum squeezing~\cite{scully1997quantum} in the FMR mode.
Here, the choice of the variables to quantize is dictated by the exact form of the relation~\eqref{eqMyMxCommutator}.

To proceed, we need to determine the magnetic free energy of the beam.

\subsection{Magnetic free energy}

We take into account the Zeeman free energy and the free energy due to the demagnetization and magnetoelasticity.
The latter two provide the coupling to the mechanical mode of the beam.
We neglect the crystal anisotropies.
That can be a reasonable approximation for a polycrystalline ferromagnetic material, for example CoFeB~\cite{kansanen2021magnomechanics}.
Also, we assume that the shape anisotropy in the plane of the beam is weak.

We provide the general expressions for the free energies of the beam.
The Zeeman energy
\begin{equation}
\label{eqFZeemanGeneral}
	F_{\Ze} = -\mu_0 \int dV \vec H(t) \cdot \vec M
\end{equation}
is responsible for aligning the magnetization $\vec M$ along the direction of the external magnetic field $\vec H$.
Here and below $\int dV$ stands for the integration over the beam volume.

\begin{figure}
\includegraphics{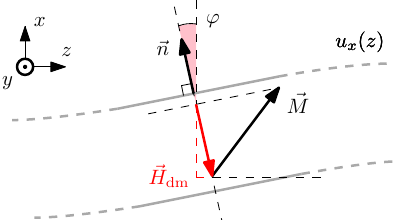}
\caption{Demagnetizing field $\vec H_{\dm}$ in a small section of the beam.
The beam is displaced by $u_x(z)$ at each point $z$.
Its magnetization is $\vec M$.
The section normal vector $\vec n$ is tilted by angle $\varphi$ with respect to the $x$-axis.}
\label{figDemagnetization}
\end{figure}

The demagnetizing energy reads
\begin{equation}
\label{eqFDemagnetizingGeneralPlate}
	F_{\dm} = -\mu_0 \int dV \vec H_{\dm} \cdot \vec M.
\end{equation}
Here $\vec n = -\vec e_z \sin\varphi + \vec e_x \cos\varphi$ denotes the beam normal vector.
The demagnetizing field $\vec H_{\dm} = - (\vec M \cdot \vec n) \, \vec n$ compensates the normal component of the magnetization due to the surface currents arising.
Its direction is set by the rotational symmetry of a thin plate.
One can obtain the demagnetizing field from the continuity of the magnetic field $\vec B$ normal component and the magnetizing field $\vec H$ tangential component at the surface.
In the simple geometry of Fig.~\ref{figDemagnetization} we arrive at  $H_{\dm\, x} = -|\vec H_{\dm}| \cos\varphi = M_z \sin\varphi \cos\varphi - M_x \cos^2 \varphi$ and $H_{\dm\, y} = |\vec H_{\dm}| \sin\varphi = -M_z \sin\varphi + M_x \sin\varphi \cos\varphi$ for the $x$ and $y$ components of the demagnetizing field.
Plugging them into Eq.~\eqref{eqFDemagnetizingGeneralPlate} yields
\begin{equation}
\label{eqFDemagnetizing}
	F_{\dm} = \frac 1 2 \mu_0 \Big[V M_x^2
			  + \int dV (\partial_z u_x)^2 (M_z^2 - M_x^2)\Big],
\end{equation}
where we have expanded to the second order in $\sin\varphi \approx \tan\varphi = \partial_z u_x$.
In the course of integration over the beam volume, we have taken into account the boundary conditions for the displacement $u_x$ (see Fig.~\ref{figMechanics}).

The magnetoelastic energy of isotropic media reads~\cite{gurevich1996magnetization,landau1984electrodynamics}
\begin{equation}
\label{eqFMagnetoelasticIsotropicGeneral}
	F_{\me} = \frac B {\Ms^2} \int dV M_i M_j \epsilon_{ij},
\end{equation}
where $i,j = x,y,z$ and the repeated indices are summed over.
It describes a strain arising in the magnetization direction, and vice versa.
We expand the strain tensor~\cite{landau1984elasticity}
\begin{equation}
	\epsilon_{ij} =
		\frac 1 2 \Big(\frac{\partial u_i}{\partial x_j}
		 + \frac{\partial u_j}{\partial x_i}
		 + \frac{\partial u_k}{\partial x_i} \frac{\partial u_k}{\partial x_j}
		\Big)
\end{equation}
up to the second-order terms in displacement, as they provide the cross-Kerr interaction.
The beam only deflects along the $x$-axis.
Hence the only non-vanishing components of the strain are
\begin{equation}
	\epsilon_{zz} = \frac 1 2 (\partial_z u_x)^2 + \epsilon^{(0)}_{zz},
\quad
	\epsilon_{xz} = \frac 1 2 \partial_z u_x.
\end{equation}
Here $\epsilon^{(0)}_{zz} = \partial_z u_z$ is the static strain due to the initial tension $T_0$ [see Eq.~\eqref{eqBeamVibrations}].
With tension applied, a homogeneous beam expands by the same amount at its ends, hence $u_z(0) = -u_z(L)$.
With that and the boundary conditions indicated in Fig.~\ref{figMechanics}, the integration in Eq.~\eqref{eqFMagnetoelasticIsotropicGeneral} yields
\begin{equation}
\label{eqFMagnetoelastic}
	F_{\me} = \frac B {2 \Ms^2} M_z^2 \int dV (\partial_z u_x)^2.
\end{equation}
The combination of Eqs.~\eqref{eqFDemagnetizing} and \eqref{eqFMagnetoelastic} describes the coupling between the magnetization dynamics and vibrations.

\subsection{Beam tension of magnetic origin}

Here we show that the static magnetization provides a tension to the beam.
We discuss how it influences the buckling transition.

Consider the static parts of the demagnetizing~\eqref{eqFDemagnetizing} and the magnetoelastic~\eqref{eqFMagnetoelastic} free energies that are proportional to $\int_V dV (\partial_z u_x)^2 = S \int_0^L dz (\partial_z u_x)^2$.
They are of the same form as the tension part of the mechanical Hamiltonian~\eqref{eqHMechanical}.
In other words, magnetoelastic strains affect the tension that, on the other hand, can be read off from its effect on the vibration eigenfrequency.
Besides, the strains that arise due to the changing surface anisotropy of the moving beam provide a tension-like force on it.
We combine both parts into the magnetic tension
\begin{equation}
\label{eqMagneticTension}
	T_{\mg0} = S \Big[\mu_0(M_{z0}^2 - M_{x0}^2) + \frac B{\Ms^2} M_{z0}^2 \Big],
\end{equation}
that we use in the dynamical equation of the beam~\eqref{eqBeamVibrations} and in the no-buckling condition~\eqref{eqNoBuckling}.
We denote the static magnetization components with ``0'' in the subscript.

It follows that the external magnetic field can influence the beam buckling.
In the absence of strong external magnetic field, direction of the static magnetization $\vec M_0$ is defined by the crystal anisotropies or by the shape anisotropy in the plane of the beam (which we have neglected).
Consider the case when the beam is close to the buckling transition, but Eq.~\eqref{eqNoBuckling} still holds.
Let us first assume that the magnetoelastic constant $B>0$ or $B/\Ms$ is negligible to $\mu_0$.
The magnetization can be fixed in a direction where its out of the beam component $M_{x0}$ is too small to provide enough negative tension to buckle the beam.
Then, if one applies a field that re-orients the magnetization closer to the $x$-axis direction, the beam can buckle.
Such buckling was studied in Ref.~\cite{moon1968magnetoelastic}, where it is explained as solely a demagnetization effect.
Similarly, if $B<0$ and the magnetoelastic contribution is not small, the beam can also buckle if the magnetic field is applied along the beam.

In addition, outside the buckling transition, the magnetically applied tension affects the vibration eigenfrequencies.
This thus provides a possibility to control the vibration frequencies \textit{in situ} using the magnetic field.

\subsection{Full Hamiltonian}
\label{secFullHamiltonian}

We consider the case with the static magnetizing field $\vec H_0$ oriented along the beam.
Its magnetization then follows the direction of $\vec H_0$ when $|\vec H_0|$ exceeds the coercive field.
In accordance with the approximations in Sec.~\ref{secHamiltonianStructure}, we expand the free energy up to the linear terms in $M_x/\Ms$ and $M_y/\Ms$.
Zeeman energy~\eqref{eqFZeemanGeneral} becomes
\begin{equation}
\label{eqHZeeman}
	H_{\Ze} \approx \mu_0 V \Big[\frac{H_0}{\Ms} (M_x^2 + M_y^2)
					- H_x(t) M_x - H_y(t) M_y\Big]
\end{equation}
up to a constant.
Here the first term describes a circular precession of the magnetic moment around the $z$-axis.
The other terms describe the driving due to the ac magnetic field.
We denote the dynamical part of the demagnetizing free energy by $H_{\dm} = F_{\dm} - F_{\dm0}$, where $F_{\dm0} = \frac 1 2 \mu_0 S \Ms^2 \int_0^L dz (\partial_z u_x)^2$ is the static free energy.
The dynamical part of the magnetoelastic energy $H_\me$ is defined accordingly using the static magnetoelasic energy $F_{\me0} = B S \int_0^L dz (\partial_z u_x)^2$.
We introduce the creation $m^\dag$ and annihilation $m$ operators of a magnon by a substitution
\begin{equation}
	M_y = M_{\ZPF} (m^\dag + m),
\quad
	M_x = -i M_{\ZPF} (m^\dag - m),
\end{equation}
where
\begin{equation}
\label{eqMzpf}
	M_{\ZPF} = \sqrt{\hbar\gamma\Ms / 2V}
\end{equation}
is the amplitude of the zero-point fluctuations in the magnetization.
One can check that $[m, m^\dag] = 1$.
In terms of the magnon operators, magnetic part of the full Hamiltonian reads
\begin{align}
\label{eqHmgBare}
	H_{\mg} = {}& H_{\Ze} + H_{\dm} + H_{\me},
\\
\nonumber
	H_{\Ze} = {}& \hbar\omega_M (m^\dagger m + \frac 1 2)
				       -\mu_0 M_\ZPF (m^\dag + m) H_y(t),
\\
\label{eqHZeemanSecondQuantized}
				   {}& + i\mu_0 M_\ZPF (m^\dag - m) H_x(t),
\\
	H_{\dm} = {}& -\frac 1 2 \mu_0 M_{\ZPF}^2 [4m^\dag m - (m^\dag-m)^2]
					 \, \int dV (\partial_z u_x)^2,
\\
\label{eqHmeBare}
	H_{\me} = &{} -\frac B{2\Ms^2} M_{\ZPF}^2 \, 4m^\dag m
					\, \int dV (\partial_z u_x)^2,
\end{align}
where
\begin{equation}
	\omega_M = \mu_0 \gamma H_0
\end{equation}
is the frequency of isotropic magnetic precession~\cite{gurevich1996magnetization}.
In the calculation of Hamiltonian~\eqref{eqHmgBare}--\eqref{eqHmeBare} we neglect $[M_y, M_x]$ when it is compared to $\Ms$.
As it follows from the equations above, magnetization precession interacts with the beam displacements through the demagnetizing and magnetoelastic energies.
The energy of the free circular precession in Hamiltonian~\eqref{eqHZeeman} becomes the magnon number term in the second-quantized Hamiltonian~\eqref{eqHZeemanSecondQuantized}.
Therefore, $m^\dag$ creates a magnon of a circular precession.

It is convenient to operate with the operators of magnons with elliptical precession.
We apply a squeeze transformation
\begin{equation}
	S = \exp\Big[\frac \xi 2 (m^2 - m^{\dag2})\Big]
\end{equation}
with real $\xi$ to cancel the terms proportional to $m^2$ and $m^{\dag2}$ in the Hamiltonian.
Consider the Hamiltonian of the free FMR mode of the beam $H_{\mg0} = \hbar \omega_M m^\dag m + \hbar \frac s 2 (m + m^\dag)^2$ with
\begin{equation}
	s = \mu_0 V M_{\ZPF}^2 / \hbar.
\end{equation}
It transforms to
\begin{equation}
\label{eqHmgElliptic}
	H_{\mg0} \to \omega_K m^\dag m
\end{equation}
if $\sinh 2\xi = -s / \omega_K$.
Here $\omega_K = \sqrt{\omega_M(\omega_M + 2s)}$ is the Kittel frequency of the FMR in a tangentially magnetized thin plate~\cite{gurevich1996magnetization}.
While obtaining Eq.~\eqref{eqHmgElliptic}, we have used that
\begin{equation}
	m \to S^\dag m S = m \cosh\xi - m^\dag \sinh\xi,
\end{equation}
which is a partial case of the Bogoliubov transformation used in Ref.~\cite{kansanen2021magnomechanics}.

First we provide the expressions for the rest of the magnetic Hamiltonian for an arbitrary saturation ratio $\Ms/H_0 = 2s/\omega_M$.
In the squeezed frame, the interaction Hamiltonian $H_\inter = H_\dm + H_\me$ becomes
\begin{widetext}
\begin{multline}
\nonumber
	H_\inter \to -\frac{M_\ZPF^2}2
					\bigg\{\Big[4\Big(\mu_0 + \frac B{\Ms^2}\Big)\cosh2\xi
								- \mu_0 e^{2\xi}\Big] m^\dag m
						   + \Big[2\Big(\mu_0 + \frac B{\Ms^2}\Big)\sinh2\xi
								  - \frac{\mu_0}2 e^{2\xi}\Big](m^{\dag2} + m^2)
					\bigg\}
\\
		\times x_\ZPF^2 \, (a + a^\dag)^2
\,
			S \int_0^L dz (\partial_z \upsilon)^2.
\end{multline}
The drive Hamiltonian transforms to
\begin{equation}
	H_{\FMR\dr}
		\to \mu_0 M_\ZPF [-(m^\dag + m) H_y(t) e^{-\xi}
						   + i (m^\dag - m) H_x(t) e^\xi].
\end{equation}
\end{widetext}

The expressions simplify in the case of an unsaturated magnet when $H_0 \ll \Ms$, which is often the case in practice.
In that case the Kittel frequency is approximately
\begin{equation}
	\omega_K \approx \sqrt{2s\omega_M}.
\end{equation}
The coupling Hamiltonian in the squeezed frame simplifies to
\begin{gather}
\label{eqHcrossKerr}
	H_\inter \approx \hbar \frac K 4 (m^\dag + m)^2 (a^\dag + a)^2,
\\
\begin{split}
\label{eqCrossKerrCouplingStrengthUnsaturated}
	K = -2\hbar^{-1} \sqrt{\frac{\Ms}{H_0}}
\, 
		&M_\ZPF^2 \Big(\mu_0 + \frac B {\Ms^2} \Big)
\\
		\times & \, x_\ZPF^2
			S \int_0^L dz (\partial_z \upsilon)^2,
\end{split}
\end{gather}
where we have used that $-\xi \gg 1$ and $-\sinh2\xi \approx \cosh2\xi \approx \sqrt{\Ms/4H_0}$.
The drive term becomes
\begin{equation}
\label{eqHdFMR}
	H_{\FMR\dr}
		\approx -\mu_0 M_\ZPF \sqrt[4]{\frac{\Ms}{H_0}} (m^\dag + m) H_y(t),
\end{equation}
where we have used that $-\sinh\xi \approx \cosh\xi \approx \sqrt[4]{\Ms/16H_0}$.
That is, in the unsaturated limit the FMR is only susceptible to the field component that lies along its width.
In that limit, the shape anisotropy is very strong which almost prevents the magnetization to deflect out of the beam plane.
The full system Hamiltonian now consists of the cross-Kerr Hamiltonian~\eqref{eqHcrossKerr}, the FMR and the mechanical drive Hamiltonians~\eqref{eqHdFMR} and~\eqref{eqHdVib}, and the Hamiltonians of the free FMR and mechanical modes~\eqref{eqHmgElliptic} and~\eqref{eqHFreeVibrationsSecondQuantized}.
In the main part of the paper, we assume harmonic driving.
For that, one sets the driving force $f_{\ext} \propto -\sin \Omega_\vib t$ and the ac magnetic field $H_y \propto -\sin \Omega_\FMR t$.
We also apply the RWA.
Then the Hamiltonian derived here coincides with Hamiltonian~\eqref{eqH} that we use throughout the main part of the paper, apart from the fact that the latter also includes a linear interaction term.
Also note the difference in notations.

The assumption of the unsaturated magnet often can be reasonable for CoFeB used in Ref.~\cite{kansanen2021magnomechanics}.
Its saturation magnetization is~\cite{kansanen2021magnomechanics,taniguchi2008penetration} $\Ms \approx \qty{1.2e6}{A/m}$ and its gyromagnetic ratio is~\cite{taniguchi2008penetration} $\gamma \approx \qty{1.9e11}{rad.s^{-1}.T^{-1}} \approx 2\pi \, \qty{30.2}{GHz/T}$.
With the magnetic permeability $\mu_0 \approx 4\pi \times 10^{-7}$~A/m one obtains $2s = 2\pi \times 45.6$~GHz.
With a sufficiently small static magnetizing field $H_0 \sim \Ms / 10$, the Kittel frequency $\omega_K \approx 2\pi \times 14.4$~GHz is still reasonably high.

An RWA cross-Kerr term is derived in the Supplemental Materials to Ref.~\cite{shen2022bistability} for an yttrium-iron-garnet sphere.
A modification of that derivation that does not use the RWA produces the term of the same form as in Eq.~\eqref{eqHcrossKerr}.

\subsection{Conditions for the magnetization to be homogeneous}
\label{secHomogeneousM}

Deflection of the beam from the static configuration provides an inhomogeneity for its magnetization, due to changing shape anisotropy and strain~[see Eqs.~\eqref{eqFDemagnetizing} and~\eqref{eqFMagnetoelastic}].
However, change of the magnetization direction over the beam length is penalized by the exchange interaction.
The change can be neglected if the respective energy greatly exceeds the dynamical parts of the demagnetizing and the magnetoelastic energies:
\begin{equation}
\label{eqHomogeneousMGeneral}
\int_{V/2} dV \frac{\alpha |\nabla \vec M|^2}{2\Ms^2}
	\gg \frac 1 2 \Big(2\mu_0 + \frac B{\Ms^2}\Big) M_x^2
		\int_{V/2} dV (\partial_z u_x)^2,
\end{equation}
where $\alpha$ is the exchange stiffness.
It is enough to consider a half of the beam in its first symmetric mode.
As before, we consider the case when the static magnetization part is aligned along the bridge.
Let $u_x$ denote the largest displacement of the beam at its center at $L/2$.
Suppose that the $x$ component of the magnetization changes from zero at the bridge end at $z=0$ to $M_x$ at $z = L/4$.
Then we can estimate $\partial_z u_x \sim \frac{u_x}{L/2}$ and $|\nabla M|^2 \sim (\frac {M_x}{L/4})^2$.
The condition above becomes
\begin{equation}
\label{eqHomogeneousM}
	\frac{4\alpha}{\Ms^2} \gg \Big(2\mu_0 + \frac B{\Ms^2}\Big) u_x^2.
\end{equation}
A similar condition holds for the $y$ component of the magnetization.

We can assess the values of the bridge deformation $u_x$ for which the magnetization can be considered homogeneous.
We estimate the saturation magnetization of a CoFeB beam as~\cite{kansanen2021magnomechanics,taniguchi2008penetration} $\Ms \approx \qty{1.2e6}{A/m}$.
Magnetoelastic coupling of a thin (1--2 nm) CoFeB film varies from $B = \qty{-4e6}{J/m^3}$ to $\qty{-10e6}{J/m^3}$ depending on its thickness and the preparation details~\cite{gowtham2016thickness}.
One can check that the magnetoelastic energy is then of the same order as the demagnetizing energy in Eq.~\eqref{eqHomogeneousMGeneral}.
As $B<0$, their sum ranges from a small value to the values comparable to the magnetoelastic energy alone.
In that case, to assure that the condition~\eqref{eqHomogeneousM} holds, we can compare the exchange energy to the magnetoelastic one alone.
With~\cite{choi2020exchange} $\alpha \approx \qty{13}{pJ/m}$, that yields that only with a small beam deformation $u_x \lesssim \qty1{nm}$ its magnetization can be considered homogeneous.

\subsection{Estimate of the cross-Kerr coupling strength}
\label{secEstimateK}

Here we estimate the order of magnitude of the cross-Kerr coupling strength using Eq.~\eqref{eqCrossKerrCouplingStrengthUnsaturated}.

With Eq.~\eqref{eqMzpf} one has $V M_\ZPF^2 \approx \qty{1.2e-17}{A.m}$, where $V$ denotes the beam volume, and we have used the same material parameters of CoFeB as in Sec.~\ref{secFullHamiltonian}.
Consider a bridge of the same dimensions as in Ref.~\cite{kansanen2021magnomechanics}: length $L \approx \qty{50}{\micro\meter}$, width $w \approx \qty{10}{\micro\meter}$ and height $h = \qty{50}{nm}$.
We estimate the vibration frequency $\omega_\vib \sim 2\pi \times \qty1{MHz}$ to be the same as in the reference.
Using that CoFeB mass density is~\cite{peng2016fast} \qty{8e3}{g/cm^3} one arrives at $V x_\ZPF^2 \approx \qty{1.0e-45}{m^5}$ using Eq.~\eqref{eqXinTermsOfa}.
The shape of the first vibration mode is~\cite{landau1984elasticity} $\upsilon \propto (\sin kL - \sinh kL)(\cos kz - \cosh kz) - (\cos kL - \cosh kL)(\sin kz - \sinh kz)$, where we neglect the influence of the initial and the magnetic tensions.
We find the mode normalization [see Eqs.~\eqref{eqVibModesExpansion}] and use that $k \approx 4.73/L$, which allows one to calculate $\int_0^L dz (\partial_z \upsilon)^2 \approx 0.004 / L$.
Finally, we obtain the coupling rate $-K \sim 2\pi \, \qty{10}{\pico\hertz}$, where we have taken $\sqrt{\Ms/H_0} \approx 3.3$ as in Sec.~\ref{secFullHamiltonian} and $(\mu_0 + B/\Ms^2) \sim \mu_0$ as in Sec.~\ref{secHomogeneousM}.
Note that in the high-tension limit, the mode shape is simply $\upsilon \propto \cos \pi z/L$, and $\int_0^L dz (\partial_z \upsilon)^2 = \pi^2 / L$.
We conclude that higher tension in the beam can increase $|K|$ by several orders of magnitude.

One can also reach higher cross-Kerr coupling strength with smaller width and height of the beam.
We estimate~\cite{landau1984elasticity} $\omega_\vib \sim \frac{\pi^2}{4L^2} \sqrt{\frac{EI_y}{\rho S}}$, where again we neglect the tensions.
That scales as $\omega_\vib \propto h/L^2$ as the beam cross section area is $S = wh$ and its area moment of inertia is $I_y = wh^3/12$.
One readily finds that the cross-Kerr coupling strength scales as $K \propto 1/wh^2$.
Making the beam width and height both ten times smaller than in Ref.~\cite{kansanen2021magnomechanics}, i.e.\ $w = \qty1{\micro\meter}$ and $h = \qty5{nm}$, the coupling strength is well in the nanohertz range.
Even larger cross-Kerr coupling can be obtained using atomically thin magnetic two-dimensional materials such as CrI$_3$ or CrCl$_3$ and the related compounds~\cite{huang2017layer2d,tartaglia2020accessing2d}.

\bibliography{magno,optmech,nonlin,jj_sources,cqed,counters,common_sources,comp,my}

\end{document}